\documentstyle{article}

\begin{document}

\def\levelset{{\cal X}}
\def\fxx{{\bf x}}
\def\fxy{{\bf y}}
\def\real{I\!\!R}

\begin{center}
{\large {\bf A More Rational and Perfect Scheme for C, P, T Transformations }}\\
{\large {\bf as well as C, T and CP Violations in the Regularization and }}\\
{\large {\bf Renormalization Processes of High Order Perturbations }}\\
\vskip 0.2in
{\large Mei Xiaochun} \\
\vskip 0.2in
\par
(Department of Physics, Fuzhou University, Fuzhou, 350025, China, E-mail:mxc001@163.com)
\end{center}
{\bf Abstracts } According to the current transformation theory in particle physics, after time reversal, the creation operator of a particle is still a creation operator and the annihilation operator of a particle is still an annihilation operator. This kind of definition is improper. In interaction processes, creation operator should become annihilation operator and annihilation operator should become creation operator after time reversal. There also exist some problems in the current $P$ and $C$ transformations. A more rational and perfect scheme for $C,P,T$ transformations is advanced. In new scheme, the separate $C,P,T$ transformations of transition probabilities are completely the same as the current theory when the regularizations and renormalizations of high order processes are not considered. Under the united $CPT$ transformation, the Hamiltonians of strong, weak and electromagnetic interactions are invariable with a completely symmetrical form $CPT{\cal{H}} (x)(CPT)^{-1}={\cal{H}} (x)$. On the other hand, according to new scheme, the propagation functions of Femions would change a negative sign, while the propagation functions of bosons is unchanged. In this way, when mass renormalization is considered in the high order processes, new $T,C$ violation and $CP$ violation would be caused. Meanwhile, the regularization and renormalization of the third order vertex angle processes would also violate $T,C$ and $CP$ symmetries in new scheme. But the united $CPT$ symmetry still holds in high order renormalization processes. It leads to the result that positive and negative electrons have different anomalous magnetic moments. The Compton scattering experiment and the measurements of anomalous magnetic moments of positive and negative electrons are suggested to verify the existence of $C$ violation in the high order processes. The result can be used to explain the problem of reversibility paradox in the non-equilibrium evolution processes of macro-systems, as well as the problem of positive and anti- material's asymmetry in current university. Meanwhile, in new scheme, time reversal operator is not an anti-Hermitian operator again so that the basic demand of quantum mechanics can be satisfied. The parities of transverse, longitudinal and scalar photons also become identical. So new scheme is more effective and rational with more perfect symmetry comparing with the current scheme.
\\
\\
{\bf 1.Introduction}
\par
According to the current transformation theory in particle physics, after time reversal, the creation operator of a particle is still a creation operator and the annihilation operator of a particle is still an annihilation operator. This kind of definition can't coincide with practical situations. In interaction process, particle's creation operator should become annihilation operator and annihilation operator should become creation operator after time reversal. So the current theory of time reversal should be reformed. According to new definition, particle's creation and annihilation operators exchange each after time reversal. In this way, the propagation functions of fermions change a negative sign while the propagation functions of bosons are unchanged under time reversal. For the strong and electromagnetic interactions when regularization and renormalization are not considered, the transition probability densities keep unchanged under time reversal. But in some weak interaction processes just as $K^{0}$ and $B^{0}$ meson's decays, time reversal symmetry is also violated owing to the fact that same $CKM$ matrix elements are complex with $U^{\ast}_{jk}\neq{U}_{jk}$. 
\par
It is also proved that according to new scheme, the regularization and renormalization processes of some high order perturbations would cause $C,T$ and $CP$ violations. These results can be used to explain the irreversibility paradox problem of time reversal in non-equilibrium evaluation processes of macro-systems. 
\par
On the other hand, in the current theory of $P$ transformation, the parity of transverse and longitudinal photons is defined as $-1$ but the parity of scalar photons is defined as $+1$. This is inconsistent. In the current theory of $C$ transformation, the transformation relations between spinor positive and anti-particles are defined as $\psi_c=\gamma_2\gamma_4\bar{\psi}^{\tau}$ and $\bar{\psi}_c=\psi^{\tau}\gamma_2\gamma_4$. However, in the quantum theory of field, we define $\psi=\psi^{(+)}+\psi^{(-)}$ and $\bar{\psi}=\bar{\psi}^{(+)}+\bar{\psi}^{(-)}$. The operators of second quantization $\psi$ and $\bar{\psi}$ have contained both the component of positive particles $\psi^{(-)}$ and $\bar{\psi}^{(+)}$ as well as the component of anti-particles $\psi^{(+)}$ and $\bar{\psi}^{(-)}$. It is improper to regard $\psi$ and $\bar{\psi}$ as the wave function of positive particles again. Similarly, after $C$ transformation, $\psi_c$ and $\bar{\psi}_c$ also contain the both components of positive particles and anti-particles $\psi^{(+)}$ and $\bar{\psi}^{(-)}$. It is also improper to regard $\psi_c$ and $\bar{\psi}_c$ as the wave function of anti-particles again. The real meaning of $C$ transformation in the quantum theory of field should be that in the coordinate space the creation operator $\bar{\psi}^{(+)}_s(x)$ of spinor positive particle exchange with the creation operator $\psi^{(+)}_s(x)$ of spinor anti-particle. The annihilation operator $\psi^{(-)}_s(x)$ of spinor positive particle exchanges with the annihilation operator$\bar{\psi}^{(-)}_s(x)$ of spinor anti-particle. The result leads to that the propagation function of fermions also changes a negative sign under $C$, while the propagation function of boson is unchanged. For the strong and electromagnetic interaction processes when regularization and renormalization are not considered, transition probabilities keep unchanged under new $C$ transformation. Meanwhile, in some processes of weak interaction such as $K^{0}$ and $B^0$ meson's decays, $CP$ and $T$ symmetries are violated simultaneously and both violations are just complementary. That is to say when the regularization and renormalization are not considered, the calculation results of transition probabilities are completely the same as that in the current theory under new $C,P$ transformations.
\par
However, when regularization and renormalizations are considered, new $C,T$ violation and $CP$ violation would be caused in some processes of strong, weak and electromagnetic interactions. It leads to that positive and negative electrons have different anomalous magnetic moments. The Compton scattering experiment and the measurements of anomalous magnetic moments of positive electrons are suggested to verify the existence of $C$ violation in the high order renormalization processes. The results can be used to explain the asymmetry problem of positive and anti-material in our university at present. 
\par
In this way, a more rational $C,P,T$ transformation can be achieved. The Hamiltonians of strong, weak and electromagnetic interactions are unchanged under united $CPT$ transformation with a completely symmetrical form $CPT{\cal{H}}(x)(CPT)^{-1}={\cal{H}}(x)$ in new scheme, even though the regularizations and renormalizations of high order processes are considered. Comparing with the result of current theory with $CPT{\cal{H}}(x)(CPT)^{-1}={\cal{H}}^{+}(-x)$, new scheme is more symmetrical and effective comparing with the current scheme. Meanwhile, time reversal operator is not an anti-Hermitian operator again so that the basic demand of quantum mechanics can be satisfied, and the parities of transverse, longitudinal and scalar photons also become identical in new scheme. 
\\
\\
{\bf 2. T Transformation }
\par  
In quantum mechanics, if we suppose that the Hamiltonian is invariable under time reversal with relation $TH(\vec{x},t){T}^{-1}=H(\vec{x},t)$, the form of Schrodinger's equation would be unchanged, but wave functions would become its complex conjugation with $T\psi(\vec{x},t)=\psi^{\ast}(\vec{x},t)$. It is just in this meaning, we say that the processes described by the Schrodinger's equation are unchanged under time reversal. It should be noted that in the processes described by the Schrodinger's equation, there exist no particle's creation and annihilation in general. The phenomena of particle's creation and annihilation should be described by the quantum theory of field. In the quantum theory of field, scalar field $\varphi$, electromagnetic field $A_{\mu}$ and spinor field $\psi$ are regarded as operators. There are two schemes for time reversal transformations in current particle physics. In the first scheme, the $T$ transformations of $\varphi$, $A_{\mu}$ and $\psi$ are defined individually as below
\begin{equation}
T\varphi(\vec{x},t){T}^{-1}=\varphi(\vec{x},-t)~~~~~~~~~~~~~~~~~TA_{\mu}(\vec{x},t)T^{-1}=-A_{\mu}(\vec{x},t)
\end{equation}
\begin{equation}
T\psi(\vec{x},t){T}^{-1}=\tilde{T}\psi(\vec{x},-t)=i\gamma_1\gamma_3\psi(\vec{x},-t)
\end{equation}
Where $\tilde{T}=i\gamma_1\gamma_3$ is a matrix. It is obvious that the time reversal definition shown in Eqs.(10) and (2) are different from that in quantum mechanics with $T\psi(\vec{x},t)=\psi^{\ast}(\vec{x},-t)$. Because $\varphi$, $A_{\mu}$ and $\psi$ are regarded as operators in the quantum theory of field, not to be probability amplitudes, the differences seem to be allowed. By the definitions above, it can be proved that the Hamiltonian of electromagnetic interaction
\begin{equation}
{\cal{H}}(\vec{x},t)=-{{ie}\over{2}}{A}_{\mu}(\vec{x},t)[\bar{\psi}(\vec{x},t)\gamma_{\mu}\psi(\vec{x},t)-\psi^{\tau}(\vec{x},t)\gamma^{\tau}_{\mu}\bar{\psi}^{\tau}(\vec{x},t)]
\end{equation}
satisfies following transformation relation
\begin{equation}
T{\cal{H}}(\vec{x},t){T}^{-1}={\cal{H}}(\vec{x},-t)
\end{equation}
\par
The result shows that thought the Hamiltonian of electromagnetic interaction in particle physics is not invariable, the calculation of transition probability in light of Eq.(4) in momentum space is still unchanged under time reversal. Also in this meaning, we say that the interaction processes with particle's creations and annihilations is symmetrical under time reversal.
\par
In order to satisfy Eq.(2), in the current theory, $T$ transformation of quantized spinor field $\psi(\vec{x},t)$ is carried out according to following procedure $^{(1)}$. Let $\psi(\vec{x},t)=\psi^{(-)}(\vec{x},t)+\psi^{(+)}(\vec{x},t)$ with
\begin{equation}
\psi^{(-)}(\vec{x},t)=\sum_{\vec{p},s}\sqrt{{m\over{E}}}{u}_{s}(\vec{p}){b}_{s}(\vec{p}){e}^{i(\vec{p}\cdot\vec{x}-Et)}~~~~~~~\psi^{(+)}(\vec{x},t)=\sum_{\vec{p},s}\sqrt{{m\over{E}}}{\nu}_{s}(\vec{p}){d}^{+}_{s}(\vec{p}){e}^{-i(\vec{p}\cdot\vec{x}-Et)}
\end{equation}
Because $\tilde{T}$ is only a matrix, we have
\begin{equation}
\tilde{T}\psi(\vec{x},-t)=\sum_{\vec{p},s}\sqrt{{m\over{E}}}[\tilde{T}{u}_{s}(\vec{p}){b}_{s}(\vec{p}){e}^{i(\vec{p}\cdot\vec{x}+Et)}+\tilde{T}{\nu}_{s}(\vec{p}){d}^{+}_{s}(\vec{p}){e}^{-i(\vec{p}\cdot\vec{x}+Et)}
\end{equation}
It can be proved to have following relations $^{(1)}$£º
\begin{equation}
\tilde{T}{u}_s(\vec{p})=i\gamma_1\gamma_3{u}_s(\vec{p})=\eta_{u}{u}^{\ast}_s(-\vec{p})~~~~~~~~\tilde{T}{\nu}_s(\vec{p})=i\gamma_1\gamma_3{\nu}_s(\vec{p})=\eta_{\nu}{\nu}^{\ast}_s(-\vec{p})
\end{equation}
Take $\eta_{m}=\eta_{\nu}=1$ for convenience, Eq.(6) becomes
$$\tilde{T}\psi(\vec{x},-t)=\sum_{\vec{p},s}\sqrt{{m\over{E}}}[u^{\ast}_{s}(-\vec{p}){b}_{s}(\vec{p}){e}^{i(\vec{p}\cdot\vec{x}+Et)}+\nu^{\ast}_{s}(-\vec{p}){d}^{+}_{s}(\vec{p}){e}^{-i(\vec{p}\cdot\vec{x}+Et)}]$$
\begin{equation}
=\sum_{\vec{p},s}\sqrt{{m\over{E}}}[u^{\ast}_{s}(\vec{p}){b}_{s}(-\vec{p}){e}^{-i(\vec{p}\cdot\vec{x}-Et)}+\nu^{\ast}_{s}(\vec{p}){d}^{+}_{s}(-\vec{p}){e}^{i(\vec{p}\cdot\vec{x}-Et)}]
\end{equation}
One the other hand, in the current theory, $T$ is an anti-Hermitian operator with nature. That is to say, wave functions are transformed into their complex conjugate functions after time reversal. So we have
\begin{equation}
T\psi(\vec{x},t)T^{-1}=\sum_{\vec{p},s}\sqrt{{m\over{E}}}[u^{\ast}_{s}(\vec{p}){T}b_{s}(\vec{p})T^{-1}{e}^{-i(\vec{p}\cdot\vec{x}-Et)}+\nu^{\ast}_{s}(\vec{p})T{d}^{+}_{s}(\vec{p})T^{-1}{e}^{i(\vec{p}\cdot\vec{x}-Et)}]
\end{equation}
Comparing Eq.(8) with (9), we obtain
\begin{equation}
Tb_s(\vec{p}){T}^{-1}=b_s(-\vec{p})~~~~~~~~~~~~~~~Td^{+}_s(\vec{p})T^{-1}=d^{+}_s(-\vec{p})
\end{equation}
Eq.(10) shows that creation operator is still creation operator, and annihilation operator is still annihilation operator after time reversal. Though it seams rational for free particles without creation and annihilation, it is improper for processes to exist interaction between particles. In practical interaction processes with particle's creations and annihilations, creation operator should become annihilation operator and annihilation operator should become creation operator after time reversal. So in interaction processes, the rational definitions of time reversal transformations of creation and annihilation operators should be
\begin{equation}
Tb_s(\vec{p}){T}^{-1}=b^{+}_s(-\vec{p})~~~~~~~~~~~~~~~Td^{+}_s(\vec{p})T^{-1}=d_s(-\vec{p})
\end{equation}
Let $T\psi(\vec{x},t)T^{-1}=\psi_{T}(\vec{x},t)$, $\tilde{x}=(\vec{x},-t)$, we should have in coordinate space
\begin{equation}
\psi^{(-)}_{s}(x)=\sum_{\vec{p}}\sqrt{{m\over{E}}}{u}_s(\vec{p}){b}_s(p){e}^{ip\cdot{x}}\rightarrow^{T}\bar{\psi}^{(+)}_{s}(\tilde{x})=\sum_{\vec{p}}\sqrt{{m\over{E}}}\bar{u}_s(-\vec{p}){b}^{+}_{s}(-\vec{p})e^{-ip\cdot{x}}
\end{equation}
\begin{equation}
\psi^{(+)}_{s}(x)=\sum_{\vec{p}}\sqrt{{m\over{E}}}{\nu}_s(\vec{p}){d}^{+}_s(p){e}^{-ip\cdot{x}}\rightarrow^{T}\bar{\psi}^{(-)}_{s}(\tilde{x})=\sum_{\vec{p}}\sqrt{{m\over{E}}}\bar{\nu}_s(-\vec{p}){d}_{s}(-\vec{p})e^{ip\cdot{x}}
\end{equation}
The formulas indicate that the operator $\psi^{(-)}_{s}(x)$ to annihilate a positive particle becomes the operator $\bar{\psi}^{(+)}_{s}(\tilde{x})$  to create a positive particle, and the operator $\bar{\psi}^{(+)}_{s}(x)$ to create an anti-particle becomes the operator $\psi^{(-)}_{s}(\tilde{x})$ to annihilate an anti-particle in coordinate space after time reversal. The same problems exist in the time reversals of scalar field $\varphi$, electromagnetic field $A_{\mu}$ when there exist particle's creations and annihilations. 
\par
 The same problems also exist in the second scheme of time reversal, i.e. the so-called Winger scheme $^{(2)}$, but it is unnecessary for us to discuss it any more here. As shown below, based on the correct definitions of creation and annihilation operator's time reversals, a really rational and more perfect time reversal theory can be achieved.The following points can be regarded as the foundational natures of time reversal in particle physics.
\par
1.Let $t\rightarrow-t$, $\vec{p}\rightarrow-\vec{p}$ in wave functions and other functions to describe micro-particles.
\par
2.Particle's creation and annihilation operators exchange each other. 
\par
3.The wave functions of spinor particles in momentum space are transformed into their conjugate forms. 
\par
The concrete transformations are discussed below. For free scalar field $\varphi(\vec{x},t)=Ae^{i(\vec{p}\cdot\vec{x}-Et)}$, let $t\rightarrow-t$, $\vec{p}\rightarrow-\vec{p}$ after time reversal, we have
\begin{equation}
T\varphi(\vec{x},t)=Ae^{-i(\vec{p}\cdot\vec{x}-Et)}=\varphi^{\ast}(\vec{x},t)\neq\varphi(\vec{x},-t)
\end{equation}
The wave function is changed into its complex conjugate form according to new scheme. But it is not the same as the present result shown in Eq.(1). A more serious problem is that if we only let $t\rightarrow-t$ but keep $\vec{p}$ unchanged under time reversal as done in current scheme, we have 
\begin{equation}
T\varphi(\vec{x},t)= Ae^{i(\vec{p}\cdot\vec{x}+Et)}=\varphi (\vec{x},-t)
\end{equation}
It means that retarded wave would become advanced wave. This is impossible, forbidden by the law of causation. For free and real quantized scalar field $\varphi(\vec{x},t)=\varphi^{(+)}(\vec{x},t)+\varphi^{(-)}(\vec{x},t)$
\begin{equation}
\varphi^{(+)}(x)={1\over{(2\pi)^{3/2}}}\int^{\vec{p}=+\infty}_{\vec{p}=-\infty}{{d^3\vec{p}}\over{\sqrt{2E}}}{a}^{+}(\vec{p}){e}^{-i(\vec{p}\cdot\vec{x}-Et)}~~~~~\varphi^{(-)}(x)={1\over{(2\pi)^{3/2}}}\int^{\vec{p}=+\infty}_{\vec{p}=-\infty}{{d^3\vec{p}}\over{\sqrt{2E}}}{a} (\vec{p}){e}^{i(\vec{p}\cdot\vec{x}-Et)}
 \end{equation}
we have
\begin{equation}
Ta^{+}(\vec{p})T^{-1}=\eta{a}(-\vec{p})~~~~~~~~~~~~~~~~~Ta(\vec{p})T^{-1}=\eta^{+}{a}^{+}(-\vec{p})
\end{equation}
Take $\eta=\eta^{+}=1$ for simplification and let $t\rightarrow-t$, $\vec{p}\rightarrow-\vec{p}$ after time reversal, we get 
\begin{equation}
T\varphi^{(+)}(\vec{x},t)T^{-1}=-{1\over{(2\pi)^{3/2}}}\int^{\vec{p}=-\infty}_{\vec{p}=+\infty}{{d^3\vec{p}}\over{\sqrt{2E}}}{a}(-\vec{p}){e}^{i(\vec{p}\cdot\vec{x}-Et)}
\end{equation}
Let $-\vec{p}\rightarrow\vec{p}$ again, the formula above becomes
\begin{equation}
T\varphi^{(+)}(\vec{x},t)T^{-1}={1\over{(2\pi)^{3/2}}}\int^{\vec{p}=+\infty}_{\vec{p}=-\infty}{{d^3\vec{p}}\over{\sqrt{2E}}}{a}(\vec{p}){e}^{i(-\vec{p}\cdot\vec{x}-Et)}=\varphi^{(-)}(-\vec{x},t)
\end{equation}
Similarly, we have $T\varphi^{(-)}(\vec{x},t){T}^{-1}=\varphi^{(+)}(-\vec{x},t)$. By writing $\bar{x}=(-\vec{x},t)$ in later discussions, the time reversal of quantized real scalar field is
\begin{equation}
T\varphi(x)T^{-1}=T\varphi^{(+)}(x){T}^{-1}+T\varphi^{(-)}(x)T^{-1}=\varphi^{(-)}(\bar{x})+\varphi^{(+)}(\bar{x})=\varphi(\bar{x})
\end{equation}
It seams that time reversal becomes space coordinate reversal with $\vec{x}\rightarrow-\vec{x}$. The result is equivalent to let $\vec{p}\rightarrow-\vec{p}$ at the vertex of the Feynman diagram. This just embodies the practical significance of time reversal. As shown below, it does not affect the calculation result of transition probability density. So according to this paper, when there exist particle's creations and annihilations in interaction processes, time reversal operator would not be an anti- Hermiltian one again.
\par
For complex scalar fields $\varphi(x)=\varphi^{(+)}(x)+\varphi^{(-)}(x)$, $\varphi^{+}(x)=\varphi^{+(+)}(x)+\varphi^{+(-)}(x)$ 
\begin{equation}
\varphi^{+(+)}(x)={1\over{(2\pi)^{3/2}}}\int{{d^3\vec{p}}\over{\sqrt{2E}}}{a}^{+}(\vec{p}){e}^{-ip\cdot{x}}~~~~~~~~~~\varphi^{+(-)}(x)={1\over{(2\pi)^{3/2}}}\int{{d^3\vec{p}}\over{\sqrt{2E}}}{b}(\vec{p}){e}^{ip\cdot{x}}
\end{equation}
\begin{equation}
\varphi^{(+)}(x)={1\over{(2\pi)^{3/2}}}\int{{d^3\vec{p}}\over{\sqrt{2E}}}{b}^{+}(\vec{p}){e}^{-ip\cdot{x}}~~~~~~~~~~\varphi^{(-)}(\vec{x},t)={1\over{(2\pi)^{3/2}}}\int{{d^3\vec{p}}\over{\sqrt{2E}}}{a}(\vec{p}){e}^{ip\cdot{x}}
\end{equation}
According to new scheme, we have $Ta^{+}(\vec{p}){T}^{-1}=a(-\vec{p})$, $Tb^{+}(\vec{p}){T}^{-1}=b(-\vec{p})$ and
\begin{equation}
T\varphi^{+(+)}(x){T}^{-1}=\varphi^{(-)}(\bar{x})~~~~~~~~~~~~T\varphi^{(+)}(x){T}^{-1}=\varphi^{+(-)}(\bar{x})
\end{equation}
So we get at last
\begin{equation}
T\varphi(x){T}^{-1}=\varphi^{+}(\bar{x})~~~~~~~~~~~~T\varphi^{+}(x){T}^{-1}=\varphi (\bar{x})
\end{equation}
\par
The time reversals of commutation relations between creation and annihilation operators become
\begin{equation}
T[a(\vec{p}_1),a^{+}(\vec{p}_2)]T^{-1}=[a^{+}(-\vec{p}_1),a(-\vec{p}_2)]=-[a(-\vec{p}_2),a^{+}(-\vec{p}_1)]=-\delta^3(\vec{p}_1-\vec{p}_2)
\end{equation}
\begin{equation}
T[b(\vec{p}_1),b^{+}(\vec{p}_2)]T^{-1}=[b^{+}(-\vec{p}_1),b(-\vec{p}_2)]=-[b(-\vec{p}_2),b^{+}(-\vec{p}_1)]=-\delta^3(\vec{p}_1-\vec{p}_2)
\end{equation}
Similarly, the time reversals of commutation relations between field operators become
\begin{equation}
T[\varphi^{(-)}(x_1),\varphi^{+(+)}(x_2)]T^{-1}=[\varphi^{+(+)}(\bar{x}_1),\varphi^{(-)}(\bar{x}_2)]=-[\varphi^{(-)}(\bar{x}_2),\varphi^{+(+)}(\bar{x}_1)]
\end{equation}
\begin{equation}
T[\varphi^{(+)}(x_1),\varphi^{+(-)}(x_2)]T^{-1}=[\varphi^{+(-)}(\bar{x}_1),\varphi^{(+)}(\bar{x}_2)]=-[\varphi^{(+)}(x_2),\varphi^{+(-)}(x_1)]
\end{equation}
\par
There is a difference of negative sign on the right side of the formulas comparing with the current result. In fact, it is easy to prove that the commutation relation between coordinate and momentum in quantum mechanics would change a negative sign after time reversal. From the original definition of time reversal, we can directly get $T[\hat{x},\hat{p}]T^{-1}=-[\hat{x},\hat{p}]=-i$. But in current theory of time reversal, it is artificially supposed that commutation relation between coordinate and momentum is unchanged under time reversal, i.e., $T[\hat{x},\hat{p}]T^{-1}=TiT^{-1}$, or $-[\hat{x},\hat{p}] =-i=TiT^{-1}$. From the result, time reversal operator $T$ is considered as an anti-Hermiltian operator. However, in principle, the operators of quantum mechanics should be Hermitian operators. Anti-Hermitian operators are meaningless. Therefore, it is unnecessary for us to consider time reversal operator as anti-Hermitian one generally. We should decide time reversal transformations of physical quantities in light of practical situation or original nature of time reversal processes. Though in same special cases we just have $T\alpha=\alpha^{\ast}$, but this relation has no universal meaning.
\par
The time reversal transformations of propagation functions are discussed below. The definition of propagation function of complex scalar field is
\begin{equation}
\Delta_{F}(x_1-x_2)=\theta(t_1-t_2)[\varphi^{(-)}(x_1),\varphi^{+(+)}(x_2)]+\theta(t_2-t_1)[\varphi^{+(-)}(x_2),\varphi^{(+)}(x_1)]
\end{equation}
\begin{equation}
\Delta_{F}(x_1-x_2)=-{{i}\over{(2\pi)^4}}\int^{+\infty}_{-\infty}{{d^4p}\over{p^2+m^2}}{e}^{ip\cdot(x_1-x_2)}
\end{equation}
The meaning of Eq.(29) is that when $t_1>t_2$, only the first item acts and a positive meson is created at $x_2$ point. Then the meson is annihilated when it reaches $x_1$ point. When $t_2>t_1$, only the second item acts and an anti-meson is created at point $x_1$. Then the anti-meson is annihilated when it reaches $x_2$ point. According to new scheme, by using Eq.(27) and (28), we have
\begin{equation}
T\Delta_{F}(x_1-x_2)T^{-1}=-T\theta(t_1-t_2)T^{-1}[\varphi^{(-)}(\bar{x}_2),\varphi^{+(+)}(\bar{x}_1)]-T\theta(t_2-t_1)T^{-1}[\varphi^{+(-)}(\bar{x}_1),\varphi^{(+)}(\bar{x}_2)]
\end{equation}
Because $T^{2}\sim{1}$, $T\sim\pm{1}$, we have $T\theta(t_1-t_2){T}^{-1}=\pm\theta(t_2-t_1)$. If taking $T\theta(t_1-t_2){T}^{-1}=-\theta(t_2-t_1)$, we have
\begin{equation}
T\Delta_{F}(x_1-x_2)T^{-1}=\theta(t_2-t_1)[\varphi^{(-)}(\bar{x}_2),\varphi^{+(+)}(\bar{x}_1)]+\theta(t_1-t_2)[\varphi^{+(-)}(\bar{x}_1),\varphi^{(+)}(\bar{x}_2)]
\end{equation}
It is equal to let $x_1\rightarrow\bar{x}_2$, $x_2\rightarrow\bar{x}_1$  on the right side of Eq.(29), means that when $t_2>t_1$, only the first item acts and a positive meson is create at $x_1$ point. Then the meson is annihilated when it reaches $x_2$ point. When $t_1>t_2$, only the second item acts and an anti-meson is created at point $x_2$. Then the anti-meson is annihilated when it reaches $x_1$ point. The result just coincides with the practical processes of time reversal. So let $x_1\rightarrow\bar{x}_2$, $x_2\rightarrow\bar{x}_1$ on the right side of Eq.(30), we obtain
\begin{equation}
T\Delta_{F}(x_1-x_2)T^{-1}=-{i\over{(2\pi)^4}}\int^{+\infty}_{-\infty}{{d^4{p}}\over{p^2+m^2}}{e}^{ip\cdot(\bar{x}_2-\bar{x}_1)}=\Delta_{F}(\bar{x}_2-\bar{x}_1)
\end{equation}
Because $p\cdot(\bar{x}_2-\bar{x}_1)=-\vec{p}\cdot(\vec{x}_2-\vec{x}_1)-p_0(t_2-t_1)=\vec{p}\cdot(\vec{x}_1-\vec{x}_2)+p_0(t_1-t_2)$, the formula can also be written as
\begin{equation}
T\Delta_{F}(x_1-x_2)T^{-1}=-{i\over{(2\pi)^4}}\int^{+\infty}_{-\infty}{{d^4{p}}\over{p^2+m^2}}{e}^{i[\vec{p}\cdot(\vec{x}_1-\vec{x}_2)+p_0(t_1-t_2)]}
\end{equation}
The integral above is unchanged by substitution $p_0\rightarrow-p_0$. So we can obtain Eq.(30) from Eq.(34) by substitution $p_0\rightarrow-p_0$, showing that the propagation function of scalar fields is unchanged under time reversal. In fact, we can also let $t\rightarrow-t$, $\vec{p}\rightarrow-\vec{p}$ directly on the light side of Eq.(30) to get the time reversal of propagation function of complex scalar field£¬then let $p\rightarrow-p$ again to prove the time reversal invariability of scalar field's propagation function. If taking $T\theta(t_1-t_2){T}^{-1}=\theta(t_2-t_1)$, we would get
\begin{equation}
T\Delta_{F}(x_1-x_2)T^{-1}={i\over{(2\pi)^4}}\int^{+\infty}_{-\infty}{{d^4{p}}\over{p^2+m^2}}{e}^{i[\vec{p}\cdot(\vec{x}_1-\vec{x}_2)-p_0(t_1-t_2)]}
\end{equation}
There exist the difference of a negative sign comparing with Eq.(30), means that the propagation function of scalar fields reverse its propagation direction under time reversal. This is, of course, improper. Therefore, according to the new scheme in the paper, the time reversal of step function $\theta(t)$ should be defined as
\begin{equation}
\theta(t_1-t_2)=\{^{1~~~~~~~t_1>t_2}_{0~~~~~~~t_1<t_2}\rightarrow^{T}{T}\theta(t_1-t_2){T}^{-1}=-\theta(t_2-t_1)=\{^{-1~~~~~~{t}_2>t_1}_{0~~~~~~~{t}_2<t_1}
\end{equation}
\par
For free quantized electromagnetic field $A_{\mu}(x)=\sum^{4}_{\sigma=1}{A}^{\sigma(+)}_{\mu}(x)+\sum^{4}_{\sigma=1}{A}^{\sigma(-)}_{\mu}(x)$ 
\begin{equation}
A^{\sigma(+)}_{\mu}(x)={1\over{(2\pi)^{3/2}}}\int^{+\infty}_{-\infty}{{d^3\vec{k}}\over{\sqrt{2\omega}}}\epsilon^{\sigma}_{\mu}(\vec{k}){a}^{+}_{\sigma}(\vec{k})e^{-i(\vec{k}\cdot\vec{x}-\omega{t})}
\end{equation}
\begin{equation}
A^{\sigma(-)}_{\mu}(x)={1\over{(2\pi)^{3/2}}}\int^{+\infty}_{-\infty}{{d^3\vec{k}}\over{\sqrt{2\omega}}}\epsilon^{\sigma}_{\mu}(\vec{k}){a}_{\sigma}(\vec{k})e^{i(\vec{k}\cdot\vec{x}-\omega{t})}
\end{equation}
Here $\epsilon^{\sigma}_{\mu}$ are the polarization vectors 
\begin{equation} 
\epsilon^1_{\mu}(\vec{k})=[\vec{n}_1(\vec{k}),0]~~~~~~\epsilon^2_{\mu}(\vec{k})=[\vec{n}_2(\vec{k}),0]~~~~~~\epsilon^3_{\mu}(\vec{k})=[\vec{k}/\omega,0]~~~~~~\epsilon^{4}_{\mu}(\vec{k})=[0,1]
\end{equation}
After time reversal, we have $\vec{k}\rightarrow{-\vec{k}}$, so that $\vec{n}_1\rightarrow{-\vec{n}_1}$, $\vec{n}_2\rightarrow{-\vec{n}_2}$. On the other hand, because $\epsilon^{\sigma}_{4}(\vec{k})$ represents the direction of time axis, we have $\epsilon^{\sigma}_{4}(\vec{k})\rightarrow-\epsilon^{\sigma}_4(\vec{k})$ under time reversal. So we have
\begin{equation} 
T\epsilon^1_{i}(\vec{k})T^{-1}=-\epsilon^{1}_{i}(\vec{k})~~~~~~T\epsilon^2_{i}(\vec{k})T^{-1}=-\epsilon^{2}_{i}(\vec{k})~~~~~~~T\epsilon^3_{i}(\vec{k})T^{-1}=-\epsilon^{3}_{i}(\vec{k})~~~~~~~ T\epsilon^{\sigma}_{\mu}(\vec{k})T^{-1}=-\epsilon^{\sigma}_{\mu}(\vec{k})
\end{equation}
Or
\begin{equation}
T\epsilon^{\sigma}_{\mu}(\vec{k})T^{-1}=-\epsilon^{\sigma}_{\mu}(\vec{k})
\end{equation}
According to new scheme, we also have
\begin{equation}
Ta^{+}_{\sigma}(\vec{k})T^{-1}=a_{\sigma}(-\vec{k})~~~~~~Ta_{\sigma}(\vec{k})T^{-1}=a^{+}_{\sigma}(-\vec{k})
\end{equation}
By the relations above and the same method, we get the time reversals of electromagnetic fields
\begin{equation}
TA^{\sigma(+)}_{\mu}(x)T^{-1}=-A^{\sigma(-)}_{\mu}(\bar{x})~~~~~~~~~~~~TA^{\sigma(-)}_{\mu}(x)T^{-1}=-A^{\sigma(+)}_{\mu}(\bar{x})
\end{equation}
The final result is
\begin{equation}
TA_{\mu}(x)T^{-1}=-A_{\mu}(\bar{x})
\end{equation}
Similar to scalar field, the propagation function of electromagnetic field is also unchanged under time reversal. It can also be written in new form as shown in Eq.(33) 
\begin{equation}
TD_F(x_1-x_2)_{\mu\nu}T^{-1}=-{i\over{(2\pi)^4}}\int^{+\infty}_{-\infty}{d}^4{k}{{\delta_{\mu\nu}}\over{k^2+m^2}}e^{ik\cdot(\bar{x}_2-\bar{x}_1)}=D_F(\bar{x}_2-\bar{x}_1)_{\mu\nu}
\end{equation}
\par
For free quantized spinor fields $\bar{\psi}=\sum_{s}(\bar{\psi}^{(+)}_{s}+\bar{\psi}^{(-)}_s)$ and $\psi=\sum_{s}(\psi^{(+)}_s+\psi^{(-)}_s)$,  we have
\begin{equation}
\bar{\psi}^{(+)}_{s}(x)={1\over{(2\pi)^{3/2}}}\int^{+\infty}_{-\infty}{d}^3\vec{p}\sqrt{m\over{E}}\bar{u}_s(\vec{p})b^{+}_s(\vec{p})e^{-i(\vec{p}\cdot\vec{x}-Et)}
\end{equation}
\begin{equation}
\psi^{(-)}_{s}(x)={1\over{(2\pi)^{3/2}}}\int^{+\infty}_{-\infty}{d}^3\vec{p}\sqrt{m\over{E}}{u}_{s}(\vec{p})b_s(\vec{p})e^{i(\vec{p}\cdot\vec{x}-Et)}
\end{equation}
\begin{equation}
\psi^{(+)}_{s}(x)={1\over{(2\pi)^{3/2}}}\int^{+\infty}_{-\infty}{d}^3\vec{p}\sqrt{m\over{E}}\nu_{s}(\vec{p})d^{+}_s(\vec{p})e^{-i(\vec{p}\cdot\vec{x}-Et)}
\end{equation}
\begin{equation}
\bar\psi^{(-)}_s(x)={1\over{(2\pi)^{3/2}}}\int^{+\infty}_{-\infty}{d}^3\vec{p}\sqrt{m\over{E}}\bar{\nu}_s(\vec{p})d_s(\vec{p})e^{i(\vec{p}\cdot\vec{x}-Et)}
\end{equation}
In the formula, $s=\vec{\sum}\cdot\vec{p}/\mid{\vec{p}}\mid$ is the helicity of spinor particle, $\vec{\sum}$ is particle's spin. The helicity of particle is unchanged under time reversal with $\vec{p}\rightarrow{-\vec{p}}$, $\vec{\sum}\rightarrow{-\vec{\sum}}$. So according to the new definition of time reversal or Eqs.(12) and (13), we have
\begin{equation}
Tb_s(\vec{p}){T}^{-1}=b^{+}_s(-\vec{p})~~~~~~~~~~~~Td_s(\vec{p}){T}^{-1}=d^{+}_s(-\vec{p})
\end{equation}
\begin{equation}
Tu_{s\alpha}(\vec{p})T^{-1}=\bar{u}_{s\alpha}(-\vec{p})~~~~~~~~T\nu_{s\alpha}(\vec{p})T^{-1}=\bar{\nu}_{s\alpha}(-\vec{p})
\end{equation}
Here $\alpha$ represents components. The formulas (51) can be written as the forms of matrixes
\begin{equation}
Tu_{s}(\vec{p})T^{-1}=\bar{u}^{\tau}_{s}(-\vec{p})~~~~~~~~T\nu_{s}(\vec{p})T^{-1}=\bar{\nu}^{\tau}_{s}(-\vec{p})
\end{equation}
Under time reversal $t\rightarrow{-t}$, $\vec{p}\rightarrow{-\vec{p}}$, we get
$$T\psi^{(-)}_{s\alpha}(\vec{x},t){T}^{-1}={1\over{(2\pi)^{3/2}}}\int^{+\infty}_{-\infty}{d}^3\vec{p}\sqrt{m\over{E}}\bar{u}_{s\alpha}(-\vec{p})b^{+}_{s}(-\vec{p})e^{-i(\vec{p}\cdot\vec{x}-Et)}$$
\begin{equation}
={1\over{(2\pi)^{3/2}}}\int^{+\infty}_{-\infty}{d}^3\vec{p}\sqrt{m\over{E}}\bar{u}_{s\alpha}(\vec{p})b^{+}_{s}(\vec{p})e^{-i(-\vec{p}\cdot\vec{x}-Et)}=\bar\psi^{(+)}_{s\alpha}(-\vec{x},t)
\end{equation}
or $T\psi^{(-)}_{s\alpha}(x){T}^{-1}=\bar{\psi}^{(+)}_{s\alpha}(\bar{x})$, $ T\psi^{(+)}_{s\alpha}(x){T}^{-1}=\bar{\psi}^{(-)}_{s\alpha}(\bar{x})$. We can also write them in the form of matrix
\begin{equation}
T\psi^{(-)}_s(x){T}^{-1}=\bar{\psi}^{(+)\tau}_s(\bar{x})~~~~~~~~~~T\psi^{(+)}_s(x){T}^{-1}=\bar{\psi}^{(-)\tau}_s(\bar{x})
\end{equation}
The final results are
\begin{equation}
T\bar{\psi}(x){T}^{-1}=\psi^{\tau}(\bar{x})~~~~~~~~~~~~~T\psi(x){T}^{-1}=\bar{\psi}^{\tau}(\bar{x})
\end{equation}
The commutation relations between creation and annihilation operators are£º
\begin{equation}
T\{b_s(\vec{p}_1),b^{+}_{s'}(\vec{p}_2)\}T^{-1}=\{b^{+}_{s}(-\vec{p}_1),b_{s'}(-\vec{p}_2)\}=\{b_{s'}(-\vec{p}_2),b^{+}_s(-\vec{p}_1)\}=\delta^{3}_{s,s'}(\vec{p}_1-\vec{p}_2)
\end{equation}
\begin{equation}
T\{d_{s}(\vec{p}_1),d^{+}_{s'}(\vec{p}_2)\}T^{-1}=\{d^{+}_{s}(-\vec{p}_1),d_{s'}(-\vec{p}_2)\}=\{d_{s'}(-\vec{p}_2),d^{+}_{s}(-\vec{p}_1)\}=\delta^{3}_{s,s'}(\vec{p}_1-\vec{p}_2)
\end{equation}
So the commutation relations between creation and annihilation operators are unchanged under time reversal. The commutation relations between field operators become
\begin{equation}
T\{\psi^{(-)}_{\alpha}(x_1),\bar{\psi}^{(+)}_{\beta}(x_2)\}T^{-1}=\{\bar{\psi}^{(+)}_{\alpha}(\bar{x}_1),\psi^{(-)}_{\beta}(\bar{x}_2)\}=\{\psi^{(-)}_{\beta}(\bar{x}_2),\bar{\psi}^{(+)}_{\alpha}(\bar{x}_1)\}
\end{equation}
\begin{equation}
T\{\psi^{(+)}_{\alpha}(x_1),\bar{\psi}^{(-)}_{\beta}(x_2)\}T^{-1}=\{\bar{\psi}^{(-)}_{\alpha}(\bar{x}_1),\psi^{(+)}_{\beta}(\bar{x}_{2})\}=\{\psi^{(+)}_{\beta}(\bar{x}_2),\bar{\psi}^{(-)}_{\alpha}(\bar{x}_1)\}
\end{equation}
\par
The time reversal of propagation function of spinor fields is discussed below. The definition of propagation function of spinor field is
\begin{equation}
S_F(x_1-x_2)_{\alpha\beta}=\theta(t_1-t_2)\{\psi^{(-)}_{\alpha}(x_1),\bar{\psi}^{(+)}_{\beta}(x_2)\}-\theta(t_2-t_1)\{\bar{\psi}^{(-)}_{\beta}(x_2),\psi^{(+)}_{\alpha}(x_1)\}
\end{equation}
Or
\begin{equation}
S_F(x_1-x_2)_{\alpha\beta}=-{i\over{(2\pi)^4}}\int^{+\infty}_{-\infty}{d}^4{p}({{m-i\hat{p}}\over{p^2+m^2}})_{\alpha\beta}{e}^{ip\cdot(x_1-x_2)}
\end{equation}
The physical meaning of Eq.(60) is that when $t_1>t_2$, only the first item acts and a spinor positive particle is created at $x_2$ point. Then the particle is annihilated when it reaches $x_1$ point. When $t_2>t_1$, only the second item acts and a spinor anti-particle is created at point . Then the anti-particle is annihilated when it reaches $x_2$ point. By using Eqs.(36), (58) and (59), the time reversal of propagation function of spinor field is
\begin{equation}
TS_F(x_1-x_2)_{\alpha\beta}T^{-1}=-\theta(t_2-t_1)\{\psi^{(-)}_{\beta}(\bar{x}_2),\bar{\psi}^{(+)}_{\alpha}(\bar{x}_1)\}+\theta(t_1-t_2)\{\bar{\psi}^{(-)}_{\alpha}(\bar{x}_1),\psi^{(+)}_{\beta}(\bar{x}_2)\}
\end{equation}
It is equivalent to let $x_1\rightarrow{\bar{x}_2}$, $x_2\rightarrow{\bar{x}_1}$, $\alpha\leftrightarrow\beta$ on the right side of Eq.(59), besides the difference of a negative sign. So we can get directly
\begin{equation}
TS_F(x_1-x_2)_{\alpha\beta}{T}^{-1}={i\over{(2\pi)^4}}\int^{+\infty}_{-\infty}{d}^4{p}
({{m-i\hat{p}}\over{p^2+m^2}})_{\beta\alpha}{e}^{ip\cdot(\bar{x}_1-\bar{x}_2)}=-S_F(\bar{x}_2-\bar{x}_1)_{\beta\alpha}
\end{equation}
\par
Let's now discuss the concrete form of operator $T$ in new scheme. Because we construct the interaction Hamiltonian in interaction representation, so it is enough for us to discuss the motion equations of free particles. By acting operator  $T$ on the motion equation of positive particle in momentum space, and considering relations $T\vec{p}{T}^{-1}=-\vec{p}$, $Tp_4{T}^{-1}=p_4$, we can get 
\begin{equation}
T(i\hat{p}+m){u}_s(\vec{p}){T}^{-1}=(-iT\vec{\gamma}{T}^{-1}\cdot\vec{p}+iT\gamma_4{T}^{-1}{p}_4+m)Tu_s(\vec{p})T^{-1}=0
\end{equation}
On the other hand, as we known that the wave function of spinor particle in momentum space is the eigen function of particle's helicity operator. Particle's helicity are unchanged when the directions of particle's spin and momentum are reversed simultaneously. So the wave function's forms of spinor particles in momentum space are unchanged when particle's helicity is invariable $^{(3)}$, i.e., we have 
\begin{equation}
\bar{u}_s(-\vec{p})=\bar{u}_s(\vec{p})~~~~~~~u_s(-\vec{p})=u_s(\vec{p})~~~~~~~\bar{\nu}_s(-\vec{p})=\bar{\nu}_s(\vec{p})~~~~~~~\nu_s(-\vec{p})=\nu_s(\vec{p})
\end{equation}
Therefore, in the processes of interaction, the time reversals of spinor particle's wave functions in momentum space can be written as 
\begin{equation}
Tu_s(\vec{p}){T}^{-1}=\bar{u}^{\tau}_s(-\vec{p})=\bar{u}^{\tau}_s(\vec{p})~~~~~~~~~~~~~~T\bar{u}_s(\vec{p}){T}^{-1}=u^{\tau}_s(-\vec{p})=u^{\tau}_s(\vec{p})
\end{equation}
\begin{equation}
T\nu_s(\vec{p}){T}^{-1}=\bar{\nu}^{\tau}_s(-\vec{p})=\bar{\nu}^{\tau}_s(\vec{p})~~~~~~~~~~~~~~T\bar{\nu}_s(\vec{p}){T}^{-1}=\nu^{\tau}_s(-\vec{p})=\nu^{\tau}_s(\vec{p})
\end{equation}
Thus, Eq.(63) can be written as
\begin{equation}
\bar{u}_s(\vec{p})[-i(T\vec{\gamma}{T}^{-1})^{\tau}\cdot\vec{p}+i(T\gamma_4{T}^{-1})^{\tau}{p}_4+m]=0
\end{equation}
Because the motion equation that $\bar{u}_s(\vec{p})$ satisfies is $\bar{u}_s(\vec{p})(i\vec{\gamma}\cdot\vec{p}+i\gamma_4{p}_4+m)=0$. Comparing with Eq.(68), we get
\begin{equation}
(T\vec{\gamma}{T}^{-1})^{\tau}=-\vec{\gamma}~~~~~~~~~~~~(T\gamma_4{T}^{-1})^{\tau}=\gamma_4
\end{equation}
We can take
\begin{equation}
T=\pm{i}\gamma_1\gamma_3\gamma_4
\end{equation}
and write
\begin{equation}
\gamma'_{\mu}=T\gamma_{\mu}{T}^{-1}=(\gamma_1,-\gamma_2,\gamma_3,\gamma_4)~~~~~\bar{\gamma}_{\mu}=\gamma^{\tau}_{\mu}=(-\gamma_1,-\gamma_2,-\gamma_3,\gamma_4)=(-\vec{\gamma},\gamma_4)
\end{equation}
\par
Therefore, the motion equations of spinor fields can't keep unchanged under time reversal according to the new scheme when there exist interaction and particle's creations and annihilations. In this case, wave functions in momentum space are transformed into their conjugate forms. The motion equations also become their conjugate forms correspondingly. This is different from free particles and free particle's motion equations. The fact that the motion equation is unchanged means that probability amplitude is unchanged. But probability amplitude can't be measured directly. What can be done directly is probability density. As shown below, it will be proved that though the interaction Hamiltonian and probability amplitudes can't keep unchanged under new time reversal, transition probability densities are still invariable.
\par
The Hamiltonian density of electromagnetic interaction can be written as
\begin{equation}
{\cal{H}}(x)=-A_{\mu}(x){J}_{\mu}(x)~~~~~~J_{\mu}(x)={{ie}\over{2}}[\bar{\psi}(x)\gamma_{\mu}\psi(x)-\psi^{\tau}(x)\gamma^{\tau}_{\mu}\bar{\psi}^{\tau}(x)]
\end{equation}
Because at the same space-time point $x$ we have 
\begin{equation}
\{\psi_{\alpha}(x),\bar{\psi}_{\beta}(x)\}=0~~~~~~~~~~~\psi_{\alpha}(x)\bar{\psi}_{\beta}(x)=-\bar{\psi}_{\beta}(x)\psi_{\alpha}(x)
\end{equation}
So we get
$$TJ_{\mu}(x)T^{-1}={{ie}\over{2}}[\psi_{\alpha}(\bar{x})(\gamma'_{\mu})_{\alpha\beta}\bar{\psi}_{\beta}(\bar{x})-\bar{\psi}^{\tau}_{\alpha}(\bar{x})(\gamma'^{\tau}_{\mu})_{\alpha\beta}\psi^{\tau}_{\beta}(\bar{x})]$$
$$=-{{ie}\over{2}}[\bar{\psi}_{\beta}(\bar{x})(\gamma'^{\tau}_{\mu})_{\beta\alpha}\psi_{\alpha}(\bar{x})-\psi^{\tau}_{\beta}(\bar{x})(\gamma'_{\mu})_{\beta\alpha}\bar{\psi}^{\tau}_{\alpha}(\bar{x})]$$
\begin{equation}
=-{{ie}\over{2}}[\bar{\psi}(\bar{x})\bar{\gamma}_{\mu}\psi(\bar{x})-\psi^{\tau}(\bar{x})\bar{\gamma}^{\tau}_{\mu}\bar{\psi}^{\tau}(x)]
\end{equation}
By means of Eqs.(44) and (72), the time reversal of interaction Hamiltonian is
\begin{equation}
T{\cal{H}}(x)T^{-1}=-{{ie}\over{2}}A_{\mu}(\bar{x})[\bar{\psi}(\bar{x})\bar{\gamma}_{\mu}\psi(\bar{x})-\psi^{\tau}(\bar{x})\bar{\gamma}^{\tau}_{\mu}\bar{\psi}^{\tau}(\bar{x})]
\end{equation}
The Hamiltonian can't keep unchanged under the time reversal with the difference $\gamma_{\mu}\rightarrow\bar{\gamma}_{\mu}$ and $\vec{x}\rightarrow{-\vec{x}}$. The difference $\vec{x}\rightarrow{-\vec{x}}$ does not affect transition probability. By substituting $\gamma_{\mu}$ with $\bar{\gamma}_{\mu}$ in the Feynman diagrams in momentum space, we can obtain $S$ matrixes of time reversal processes directly. The result is equivalent to reverse the momentum directions of all particle's in time reversal processes. On the other hand, taking the complex conjugation of flow $J_{\mu}(x)$, and considering the relation $\gamma_4\gamma_{\mu}=\bar{\gamma}_{\mu}\gamma_4$, we can get
\begin{equation}
J^{+}_{\mu}=-{{ie}\over{2}}[\bar{\psi}(x)\bar{\gamma}_{\mu}\psi(x)-\psi^{\tau}(x)\bar{\gamma}^{\tau}_{\mu}\bar{\psi}^{\tau}(x)]
\end{equation}
So the time reversals of the flow and the electromagnetic interaction Hamiltonian can be written as
\begin{equation}
TJ_{\mu}(x)T^{-1}=J^{+}_{\mu}(\bar{x})~~~~~~~~~T{\cal{H}}(x)T^{-1}=A_{\mu}(\bar{x})J^{+}_{\mu}(\bar{x})
\end{equation}
It will be proved blow that though the electromagnetic interaction Hamiltonian can not keep unchanged under new time reversal, the transition probabilities would invariable if the renormalization effects of high order processes are not considered. The symmetry violation of time reversal is only relative to the regularization and renormalization processes of high order processes. 
\par
The low order processes of electromagnetic interaction are discussed at first. We can discuss them in momentum space directly. For the second order process of electron-electron scattering which contains internal photon line
\par
$$e^{-}~~~~~+~~~~~e^{-}~~~=~~~e^{-}~~~~~+~~~~~~e^{-}$$
$$(p_1,r)~~~~~~~~(q_1,s)~~~~~~~~ (p_2,r')~~~~~~~~~(q_2,s')$$
let $k=p_1-p_2$, $k'=p_1-q_2$£¬transition probability amplitude is
$$S=i\delta^4(p_1+q_1-p_2-q_2){{e^2{m}^2}\over{(2\pi)^2\sqrt{E_{p_1}{E}_{p_2}{E}_{q_1}{E}_{q_2}}}}\times$$
\begin{equation}
[\bar{u}_{r'}(\vec{p}_2)\gamma_{\mu}{u}_r(\vec{p}_1){1\over{k^2}}\bar{u}_{s'}(\vec{q}_2)\gamma_{\mu}{u}_s(\vec{q}_1)-\bar{u}_{s'}(\vec{q}_2)\gamma_{\mu}{u}_r(\vec{p}_1){1\over{k'^2}}\bar{u}_{r'}(\vec{p}_2)\gamma_{\mu}{u}_s(\vec{q}_1)]
\end{equation}
By using relations $\bar{\gamma}_{\mu}\bar{\gamma}_{\mu}=(-\vec{\gamma})\cdot(-\vec{\gamma})+\gamma_4\gamma_4=\gamma_{\mu}\gamma_{\mu}$, $\delta^3(-\sum\vec{p}_i)=\delta^3(\sum\vec{p}_i)$, and Eqs.(51), (65) and (71), let $S_T=TST^{-1}$, we get
$$S_T=i\delta^{3}(-\vec{p}_1-\vec{q}_1+\vec{p}_2+\vec{q}_2)\delta(p_{10}+q_{10}-p_{20}-q_{20}){{e^2{m}^2}\over{(2\pi)^2\sqrt{E_{p_1}{E}_{p_2}{E}_{q_1}{E}_{q_2}}}}\times$$
$$[u_{r'}(-\vec{p}_2)\gamma'_{\mu}\bar{u}_{r}(-\vec{p}_1){1\over{k^2}}{u}_{s'}(-\vec{q}_2)\gamma'_{\mu}\bar{u}_{s}(-\vec{q}_1)-u_{s'}(-\vec{q}_2)\gamma'_{\mu}\bar{u}_{r}(-\vec{p}_1){1\over{k'^2}}{u}_{r'}(-\vec{p_2})\bar{\gamma}_{\mu}\bar{u}_{s}(-\vec{q}_{1})]$$
$$=i\delta^4(p_1+q_1-p_2-q_2){{e^2{m}^2}\over{(2\pi)^2\sqrt{E_{p_1}{E}_{p_2}{E}_{q_1}{E}_{q_2}}}}\times$$
\begin{equation}
[\bar{u}_r(\vec{p}_1)\gamma_{\mu}{u}_{r'}(\vec{p}_2){1\over{k^2}}\bar{u}_s(\vec{q}_1)\gamma_{\mu}{u}_{s'}(\vec{q}_2)-\bar{u}_r(\vec{p}_1)\gamma_{\mu}{u}_{s'}(\vec{q}_2){1\over{k'^2}}\bar{u}_s(\vec{q}_1)\gamma_{\mu}{u}_{r'}(\vec{p}_2)]
\end{equation}
Therefore, we have $S_T\neq{S}$. It is obvious that under time reversal the process to annihilate electrons with momentums $\vec{p}_1$ and $\vec{q}_1$ and to produce electrons with momentums $\vec{p}_2$ and $\vec{q}_2$ become the processes to produce electrons with momentums $\vec{p}_1$ and $\vec{q}_1$ and to annihilate electrons with momentums $\vec{p}_2$ and $\vec{q}_2$. But according to the current scheme, the transition probability amplitude is unchanged under time reversal with $S_T=S$. The order of process is also unchanged. So it is obvious that the current theory can't describe the practical situation of time reversal. Meanwhile, by the relation $\bar{\gamma}_{\mu}\bar{\gamma}_{\mu}=\gamma_{\mu}\gamma_{\mu}$ and $\gamma_4\gamma_{\mu}\gamma_4=\bar{\gamma}_{\mu}$, the complex conjugation of Eq.(79) is
$$S^{+}=-i\delta^{4}(p_1+q_1-p_2-q_2){{e^2{m}^2}\over{(2\pi)^2\sqrt{E_{p_1}{E}_{p_2}{E}_{q_1}{E}_{q_2}}}}\times$$
\begin{equation}
[\bar{u}_r(\vec{p}_1)\gamma_{\mu}{u}_{r'}(\vec{p}_2){1\over{k^2}}\bar{u}_{s}(\vec{q}_{1})\gamma_{\mu}{u}_{s'}(\vec{q}_2)-\bar{u}_r(\vec{p}_{1})\gamma_{\mu}{u}_{s'}(\vec{q}_2){1\over{k'^{2}}}\bar{u}_{s}(\vec{q}_1)\gamma_{\mu}{u}_{r'}(\vec{p}_{2})]
\end{equation}
So we have $S_T=-S^{+}$ and $S^{+}_T{S}_T=S^{+}{S}$, that is to say, though the transition probability amplitude can't keep unchanged under time reversal in light of the new scheme, the transition probability density is still unchanged. For the Compton scattering process which contains internal electron line
$$e^{-}~~~~+~~~~~\gamma~~~=~~~e^{-}~~~~+~~~~\gamma$$
$$(p_1,r)~~~~~~(k_1,\sigma)~~~~~~(p_2,s)~~~~~~(k_2,\tau)$$
transition probability amplitude is
\begin{equation}
S\sim{i}\bar{u}_s(\vec{p}_2)[\epsilon^{\rho}_{\nu}(\vec{k}_2)\gamma_{\nu}{{m-i\hat{p}}\over{p^2+m^2}}\gamma_{\mu}\epsilon^{\sigma}_{\nu}(\vec{k}_1)+\epsilon^{\sigma}_{\nu}(\vec{k}_1)\gamma_{\nu}{{m-i\hat{p}'}\over{p'^2+m^2}}\gamma_{\mu}\epsilon^{\rho}_{\mu}(\vec{k}_2)]{u}_r(\vec{p}_1)
\end{equation}
Here $p=p_1+k_1$, $p'=p_1-k_2$. According to new scheme and by means of Eqs.(51),£¨65£© and (71), time reversal of Eq.(81) is 
$$S_T\sim{-iu_s}(-\vec{p}_2)[\epsilon^{\rho}_{\nu}(\vec{k}_2)\gamma'_{\nu}({{m-i\hat{p}}\over{p^2+m^2}})^{\tau}\gamma'_{\mu}\epsilon^{\sigma}_{\mu}(\vec{k}_1)+\epsilon^{\sigma}_{\nu}(\vec{k}_1)\gamma'_{\nu}({{m-i\hat{p}'}\over{p'^2+m^2}})^{\tau}\gamma'_{\mu}\epsilon^{\rho}_{\mu}(\vec{k}_2)]\bar{u}_r(-\vec{p}_{1})$$
\begin{equation}
=-i\bar{u}_r(\vec{p}_1)[\epsilon^{\sigma}_{\mu}(\vec{k}_1)\bar{\gamma}_{\mu}{{m-i\hat{p}}\over{p^2+m^2}}\bar{\gamma}_{\nu}\epsilon^{\rho}_{\nu}(\vec{k}_2)+\epsilon^{\rho}_{\mu}(\vec{k}_2)\bar{\gamma}_{\mu}{{m-i\hat{p}'}\over{p'^2+m^2}}\bar{\gamma}_{\nu}\epsilon^{\sigma}_{\nu}(\vec{k}_1)]{u}_s(\vec{p}_2)
\end{equation}
So $S_T\neq{S}$. On the other hand, we have
\begin{equation}
(i\hat{p})^{+}\gamma_{4}=-i(\vec{\gamma}\cdot\vec{p}+i\gamma_4{p}_0)^{+}\gamma_4=-i(\vec{\gamma}\cdot\vec{p}-i\gamma_4{p}_0)\gamma_4=\gamma_4(i\hat{p})
\end{equation}
So the complex conjugation of Eq(81) is
\begin{equation}
S^{+}\sim-i\bar{u}_r(\vec{p}_1)[\epsilon^{\sigma}_{\mu}(\vec{k}_{1})\bar{\gamma}_{\mu}{{m-i\hat{p}}\over{p^2+m^2}}\bar{\gamma}_{\nu}\epsilon^{\rho}_{\nu}(\vec{k}_2)+\epsilon^{\rho}_{\mu}(\vec{k}_2)\bar{\gamma}_{\mu}{{m-i\hat{p}'}\over{p'^2+m^2}}\bar{\gamma}_{\nu}\epsilon^{\sigma}_{\nu}(\vec{k}_1)]{u}_s(\vec{p}_{2})
\end{equation}
We have $S_T=S^{+}$ and $S^{+}_{T}{S}_{T}=SS^{+}$. The transition probability density is still unchanged under time reversal. As for the high-order processes, by the same method, the transition probability amplitudes of electron self-energy before and after time reversal are 
\begin{equation}
S\sim\int{d}^4{k}\bar{u}_r(\vec{p})\gamma_{\mu}{{m-i(\hat{p}-\hat{k})}\over{(p-k)^2+m^2}}\gamma_{\mu}{1\over{k^2}}{u}_s(\vec{q})
\end{equation}
\begin{equation}
S_T\sim-\int{d}^{4}{k}\bar{u}_s(\vec{q}){1\over{k^2}}\gamma_{\mu}{{m-i(\hat{p}-\hat{k})}\over{(p-k)^2+m^2}}\gamma_{\mu}{u}_r(\vec{p})\sim-S^{+}
\end{equation}
The probability density is still unchanged with $S^{+}_{T}{S}_T=SS^{+}$. For simplification, imaginary number $i$ and its complex conjugation have not been written out in the formulas. For vacuum polarization process, probability amplitudes before and after time reversal are
\begin{equation}
S\sim\int\int{d}^{4}{p}d^{4}qTr[\epsilon^{\rho}_{\nu}(\vec{l})\gamma_{\nu}{{m-i\hat{p}}\over{p^2+m^2}}\gamma_{\mu}{{m-i\hat{p}}\over{q^2+m^2}}\epsilon^{\sigma}_{\mu}(\vec{k})]
\end{equation}
\begin{equation}
S_T\sim\int\int{d}^{4}p{p}d^{4}qTr[\epsilon^{\sigma}_{\mu}(\vec{k}){{m-i\hat{p}}\over{q^2+m^2}}\bar{\gamma}_{\mu}{{m-i\hat{p}}\over{p^2+m^2}}\bar{\gamma}_{\nu}\epsilon^{\rho}_{\nu}(\vec{l})]\sim{S}^{+}
\end{equation}
The probability density is also unchanged. By considering Eqs.(62) and (41), the probability amplitude of vertex process is 
\begin{equation}
S\sim\int{d}^{4}k'\bar{u}_r(\vec{p})\gamma_{\mu}{{m-i(\hat{p}-\hat{k}')}\over{(p-k')^{2}+m^{2}}}\gamma_{\nu}{{m-i(\hat{q}-\hat{k}')}\over{(q-k')^{2}+m^2}}\gamma_{\mu}{1\over{k'^2}}{u}_s(\vec{q})\epsilon^{\sigma}_{\nu}(\vec{k})
\end{equation}
\begin{equation}
S\sim-\int{d}^{4}k'\epsilon^{\sigma}_{\nu}(\vec{k})\bar{u}_{s}(\vec{q}){1\over{k'^2}}\gamma_{\mu}{{m-i(\hat{q}-\hat{k}')}\over{(q-k')^2+m^2}}\bar{\gamma}_{\nu}{{m-i(\hat{p}-\hat{k}')}\over{(p-k')^{2}+m^2}}\gamma_{\mu}{u}_r(\vec{p})\sim-S^{+}
\end{equation}
The probability density is also unchanged.
\par
For more complex high-order processes, we take following figures as an example. InFig.1, b and c can be derived from a. Let $R(\hat{p})$ and $D(k)$ represent the propagation lines of electron and photon, we get
\begin{equation}
S_1\sim\bar{u}_s(\vec{p}_2)\epsilon^{\rho}_{\nu}(\vec{k}_2)\gamma_{\nu}{R}(\hat{p})\gamma_{\mu}\epsilon^{\sigma}_{\mu}(\vec{k}_1){u}_r(\vec{p}_{1})
\end{equation}
\begin{equation}
S_2\sim\bar{u}_s(\vec{p}_2)\epsilon^{\rho}_{\nu}(\vec{k}_2)\gamma_{\alpha}{R}(\hat{p})\gamma_{\nu}{R}(\hat{p}')\gamma_{\alpha}{R}(\hat{p}^{''})\gamma_{\mu}{D}(k)\epsilon^{\sigma}_{\mu}(\vec{k}_1){u}_r(\vec{p}_{1})
\end{equation}
\begin{equation}
S_3\sim\bar{u}_s(\vec{p}_2)\epsilon^{\rho}_{\nu}(\vec{k}_2)\gamma_{\nu}{R}(\hat{p})\gamma_{\alpha}{D}(k'){R}(\hat{p}'^{''})\gamma_{\alpha}{R}(\hat{p})\gamma_{\mu}\epsilon^{\tau}_{\mu}(\vec{k}_1){u}_r(\vec{p}_{1})
\end{equation}
Because the propagation line of electron changes a negative sign but the propagation line of photon dose not after time reversal, we have 
\begin{equation}
S_{1T}\sim-\bar{u}_r(\vec{p}_1)\epsilon^{\sigma}_{\mu}(\vec{k}_1)\bar{\gamma}_{\mu}{R}(\hat{p})\bar{\gamma}_{\nu}\epsilon^{\rho}_{\nu}(\vec{k}_2){u}_s(\vec{p}_2)\sim-S^{+}_1
\end{equation}
\begin{equation}
S_{2T}\sim-\bar{u}_r(\vec{p}_{1})\epsilon^{\sigma}_{\mu}(\vec{k}_1){D}(k)\bar{\gamma}_{\mu}{R}(\hat{p}^{''})\gamma_{\alpha}{R}(\hat{p}')\bar{\gamma}_{\nu}{R}(\hat{p})\gamma_{\alpha}\epsilon^{\rho}_{\nu}(\vec{k}_2){u}_s(\vec{p}_2)\sim-S^{+}_{2}
\end{equation}
\begin{equation}
S_{3T}\sim-\bar{u}_r(\vec{p}_1)\epsilon^{\sigma}_{\mu}(\vec{k}_1)\bar{\gamma}_{\mu}{R}(\hat{p})\gamma_{\alpha}{R}(\hat{p}^{'''}){D}(k')\gamma_{\alpha}{R}(\hat{p})\bar{\gamma}_{\nu}\epsilon^{\rho}_{\nu}(\vec{k}_2){u}_s(\vec{p}_2)\sim-S^{+}_3
\end{equation}
So for the total process, we have $S=S_1+S_2+S_3$, $S_T=-(S^{+}_1+S^{+}_2+S^{+}_3)$ and $S^{+}_{T}{S}_T=S^{+}{S}$, i.e., the probability density is still unchanged under time reversal. But in Section 6 we will prove that the regularization and normalization processes of high order perturbations would cause $T$ violations, no matter in current or in new transformation schemes. 
\\
\\
\\
\\
\\
\\
\\
\\
\\
\par
Fig. 1. A complex process containing a second order and two fourth order processes
\\
\par
It can been seen from the discuss above that because the propagations of Fermion lines always continuous, the number of Fermion propagation lines in a high order diagram which is deduced from a low order diagram is always odd or always even when mass renormalization effect is not considered. So after time reversal, the total probability amplitude can always be written as $S_T=\sum{S}_{iT}=\pm\sum{S}^{+}_i$ and we always have $S^{+}_{T}{S}_T=SS^{+}$.. In this way, the transition probability densities are always unchanged under new time reversal, no matter how complex the high order processes are. But if mass renormalization is considered in the high order perturbation processes, new time reversal symmetry violation would be caused in new scheme. This problem will also be discussed in Section 6.
\par
Time reversal transformations of electro-weak interaction and strong interaction are discussed below. For electro-weak interaction between leptons, the Hamiltonian is 
\begin{equation}
{\cal{H}}(x)={\cal{H}}_e(x)+{\cal{H}}_w(x)+{\cal{H}}_z(x)
\end{equation}
Where ${\cal{H}}_e(x)$ is the Hamiltonian of electromagnetic interaction, ${\cal{H}}_w(x)$ and ${\cal{H}}_z(x)$ are the Hamiltonians of weak interaction with
\begin{equation}
{\cal{H}}_w(x)=-W^{+}_{\mu}(x){J}_{\mu+}(x)-W^{-}_{\mu}(x){J}_{\mu-}(x) \end{equation}
\begin{equation}
{\cal{H}}_z(x)=-W^{0}_{\mu}(x){J}_{\mu{0}}(x)
\end{equation}
\begin{equation}
J_{\mu{+}}(x)=i{g\over{2\sqrt{2}}}\bar{\psi}_{\nu}(x)\gamma_{\mu}(1+\gamma_5)\psi_l(x)~~~~~~~J_{\mu{-}}(x)=i{g\over{2\sqrt{2}}}\bar{\psi}_l(x)\gamma_{\mu}(1+\gamma_5)\psi_{\nu}(x)
\end{equation}
\begin{equation}
J_{\mu{0}}(x)=i{{\sqrt{g^2+g'^2}}\over{4}}[\bar{\psi}_{\nu}(x)\gamma_{\mu}(1+\gamma_5)\psi_{\nu}(x)+\bar{\psi}_l(x)\gamma_{\mu}(4{\sin}^2\theta-1-\gamma_5)\psi_l(x)]
\end{equation}
By Eq.(70), we have $T\gamma_5{T}^{-1}=-\gamma_5$ and get
$$TJ_{\mu{+}}(x){T}^{-1}=i{g\over{2\sqrt{2}}}{T}\bar{\psi}_{\nu}(x)(1-\gamma_5)\gamma_{\mu}\psi_l(x){T}^{-1}=i{g\over{2\sqrt{2}}}\psi_{\nu\alpha}(\bar{x})(1+\gamma_5)_{\alpha\beta}(\gamma'_{\mu})_{\beta\sigma}\bar{\psi}_{l\sigma}(\bar{x})$$
\begin{equation}
=-i{g\over{2\sqrt{2}}}\bar{\psi}_{l\sigma}(\bar{x})(\gamma'^{\tau}_{\mu})_{\sigma\beta}(1+\gamma^{\tau}_5)_{\beta\alpha}\psi_{\nu\alpha}(\bar{x})=-i{g\over{2\sqrt{2}}}\bar{\psi}_l(\bar{x})\bar{\gamma}_{\mu}(1+\gamma_5)\psi_{\nu}(\bar{x})
\end{equation}
Similarly
\begin{equation}
TJ_{\mu{-}}(x){T}^{-1}=-i{g\over{2\sqrt{2}}}\bar{\psi}_{\nu}(\bar{x})\bar{\gamma}_{\mu}(1+\gamma_5)\psi_l(\bar{x})
\end{equation}
\begin{equation}
TJ_{\mu{0}}(x){T}^{-1}=-i{{\sqrt{g^2+g'^2}}\over{4}}[\bar{\psi}_{\nu}(\bar{x})\bar{\gamma}_{\mu}(1+\gamma_5)\psi_{\nu}(\bar{x})+\bar{\psi}_l(\bar{x})\bar{\gamma}_{\mu}(4\sin^2\theta-1-\gamma_5)\psi_l(\bar{x})]
\end{equation}
On the other hand, taking the complex conjugation of Eqs.(100) and (101) and by means of relation $\gamma_4\gamma_{\mu}=\bar{\gamma}_{\mu}\gamma_4$, we can also get
\begin{equation}
J^{+}_{\mu{+}}(\bar{x})=TJ_{\mu{+}}(x){T}^{-1}~~~~~~~~J^{+}_{\mu{-}}(\bar{x})=TJ_{\mu{-}}(x){T}^{-1}~~~~~~~~~J^{+}_{\mu{0}}(\bar{x})=TJ_{\mu{0}}(x){T}^{-1}
\end{equation}
By defining the time reversal of gauge fields (similar to the current theory)
\begin{equation}
TW^{\pm}_{\mu}(x){T}^{-1}=-W^{\pm}_{\mu}(\bar{x})~~~~~~~~~~~~~~TZ^{0}_{\mu}(x){T}^{-1}=-Z^{0}_{\mu}(\bar{x})
\end{equation}
we have
\begin{equation}
T{\cal{H}}_w(x){T}^{-1}=W^{+}_{\mu}(\bar{x}){J}^{+}_{\mu{+}}(\bar{x})+W^{-}_{\mu}(\bar{x}){J}^{+}_{\mu{-}}(\bar{x})
\end{equation}
\begin{equation}
T{\cal{H}}_z(x){T}^{-1}=Z^0_{\mu}(\bar{x}){J}^{+}_{\mu{0}}(\bar{x})
\end{equation}
Similar to electromagnetic interaction, the differences are at $x\rightarrow\bar{x}$ and $r_{\mu}\rightarrow\bar{r}_{\mu}$ under $T$. But this kind of differences would not affect transition probabilities. To prove this point, we discuss the time reversal of propagation function of gauge fields. The propagation function of gauge fields with zero masses is
\begin{equation}
iD^{\alpha\gamma}_{\mu\nu}(x_1-x_2)=-{{i\delta_{\alpha\gamma}}\over{(2\pi)^4}}\int{d}^4{k}{{\delta_{\mu\nu}-(1-\zeta)k_{\mu}{k}_{\nu}/k^2}\over{k^2}}{e}^{ik\cdot(x_1-x_2)}
\end{equation}
According to new scheme, we have $\vec{k}\rightarrow-\vec{k}$,  $t\rightarrow-t$, under time reversal. So we can get the time reversal of Eq.(109) directly
\begin{equation}
TiD^{\alpha\gamma}_{\mu\nu}(x_1-x_2)T^{-1}=-{{i\delta_{\alpha\gamma}}\over{(2\pi)^4}}\int{d}^4{k}{{\delta_{\mu\nu}-(1-\zeta)\bar{k}_{\mu}\bar{k}_{\nu}/k^2}\over{k^{2}}}{e}^{ik\cdot(x_2-x_1)}
\end{equation}
The time reversal of propagation function of gauge fields with mass $m_w$ is 
\begin{equation}
TiD^{\alpha\gamma}_{\mu\nu}(x_1-x_2)T^{-1}=-{{i\delta_{\alpha\beta}}\over{(2\pi)^4}}\int{d}^4{k}{{\delta_{\mu\nu}+(1-1/\zeta){{\bar{k}_{\mu}\bar{k}_{\nu}}\over{k^2/\zeta+m^2_w}}}\over{k^2+m^2_w}}{e}^{ik\cdot(x_2-x_1)}
\end{equation}
The transition probability amplitude of weak interaction between four fermions in low order process, for example $\mu^{+}\rightarrow {e}^{+}+\nu_e+\tilde{\nu}_{\mu}$, can be written as (Unitary gauge is used with $\zeta\rightarrow\infty$.)
\begin{equation}
S\sim{1\over{k^2+m^2_w}}\bar{\mu}_{1r}(\vec{p}_1)\gamma_{\mu}(1+\gamma_5)\nu_{1r'}(\vec{p}^{ '}_1)(\delta_{\mu\nu}+{{k_{\mu}{k}_{\nu}}\over{m^2_w}})\bar{\nu}_{2s}(\vec{p}_2)\gamma_{\nu}(1+\gamma_5){u}_{2s'}(\vec{p}^{ '}_2)
\end{equation}
By relations $\bar{\gamma}_{\mu}\bar{\gamma}_{\mu}=\gamma_{\mu}\gamma_{\mu}$, $\bar{\gamma}_{\mu}\bar{k}_{\mu}=\gamma_{\mu}{k}_{\mu}$, transition probability amplitude under new time reversal is 
$$S_T\sim{1\over{k^2+m^2_w}}\bar{\nu}_{1r'}(\vec{p}^{ '}_1)\bar{\gamma}_{\mu}(1+\gamma_5){u}_{1r}(\vec{p}_1)(\delta_{\mu\nu}+{{\bar{k}_{\mu}\bar{k}_{\nu}}\over{m^2_w}})\bar{u}_{2s'}(\vec{p}^{ '}_2)\bar{\gamma}_{\nu}(1+\gamma_5)\nu_2(\vec{p}_{2})$$
\begin{equation}
={1\over{k^2+m^2_w}}\bar{\nu}_{1r}(\vec{p}^{ '}_1)\gamma_{\mu}(1+\gamma_5){u}_{1r}(\vec{p}_1)(\delta_{\mu\nu}+{{\bar{k}_{\mu}\bar{k}_{\nu}}\over{m^{2}_w}})\bar{u}_{2s'}(\vec{p}^{ '}_2)\gamma_{\nu}(1+\gamma_5)\nu_2(\vec{p}_2)
\end{equation}
On the other hand, by considering relation $\bar{\gamma}_{\mu}{k}^{\ast}_{\mu}=(-\vec{\gamma},\gamma_4)\cdot(\vec{k},-k_4)=-\gamma_{\mu}{k}_{\mu}$, and taking the complex conjugate of Eq.(112), we get
$$S^{+}\sim{1\over{k^2+m^2_w}}\bar{\nu}_{1r'}(\vec{p}^{ '}_1)\bar{\gamma}_{\mu}(1+\gamma_5){u}_{1r}(\vec{p}_1)(\delta_{\mu\nu}+{{\bar{k}^{\ast}_{\mu}\bar{k}^{\ast}_{\nu}}\over{m^2_w}})\bar{u}_{2s'}(\vec{p}^{ '}_2)\bar{\gamma}_{\nu}(1+\gamma_5)\nu_{2s}(\vec{p}_{2})$$
\begin{equation}
={1\over{k^2+m^2_w}}\bar{\nu}_{1r}(\vec{p}^{ '}_1)\gamma_{\mu}(1+\gamma_5){u}_{1r}(\vec{p}_1)(\delta_{\mu\nu}+{{k_{\mu}{k}_{\nu}}\over{m^{2}_w}})\bar{u}_{2s'}(\vec{p}^{ '}_2)\gamma_{\nu}(1+\gamma_5)\nu_{2s}(\vec{p}_2)
\end{equation}
Therefore, we still have  $S_T=S^{+}$ and $S^{+}_T{S}_{T}=S^{+}{S}$, transition probability density is unchanged under new time reversal. It can be proved that the results are the same in the high order processes of weak interaction between leptons without considering regularization and normalization, but we do not discuss them any more here. 
\par
For weak interaction between quarks, let $u_k=(u,c,t)$, $d_k=(d,s,b)$£¬charged currents are
\begin{equation}
J_{\mu+}(x)=i{g\over{2\sqrt{2}}}\sum^{N}_{j,k=1}{U}_{jk}\bar{u}_{j}(x)\gamma_{\mu}(1+\gamma_5){d}_k(x)
\end{equation}
\begin{equation}
J_{\mu-}(x)=i{g\over{2\sqrt{2}}}\sum^{N}_{j,k=1}{U}_{jk}\bar{d}_{j}(x)\gamma_{\mu}(1+\gamma_5){u}_{k}(x)
\end{equation}
Corresponding Hamiltonian is ${\cal{H}}(x)=-W^{+}_{\mu}(x)J_{\mu+}(x)-W^{-}_{\mu}(x)J_{\mu-}(x)$. According to new scheme, it can be thought that $U_{jk}$ has nothing to do with $T$ transformation, so we have $TU_{jk}{T}^{-1}=U_{jk}$. This is different from the current scheme. According to current theory, $T$ is an anti-Hermitian operator with transformation nature $TU_{jk}{T}^{-1}=U^{\ast}_{jk}$. Similar to Eqs.(103) and (104), we get
\begin{equation}
TJ_{\mu+}(x)T^{-1}=-i{g\over{2\sqrt{2}}}\sum^{N}_{j,k=1}{U}_{jk}\bar{d}_{k}(\bar{x})\bar{\gamma}_{\mu}(1+\gamma_5){u}_{j}(\bar{x})
\end{equation}
\begin{equation}
TJ_{\mu-}(x)T^{-1}=-i{g\over{2\sqrt{2}}}\sum^{N}_{j,k=1}{U}_{jk}\bar{u}_{k}(\bar{x})\bar{\gamma}_{\mu}(1+\gamma_5){d}_{j}(\bar{x})
\end{equation}
On the other hand, taking the complex conjugate of Eqs.(115) and (116) (Here $U_{jk}$ is only considered as a parameter, its index does not need to exchange.), we get
\begin{equation}
J^{+}_{\mu+}(x)=-i{g\over{2\sqrt{2}}}\sum^{3}_{j,k=1}{U}^{\ast}_{jk}\bar{d}_{k}(x)\bar{\gamma}_{\mu}(1+\gamma_5){u}_{j}(x)
\end{equation}
\begin{equation}
J^{+}_{\mu-}(x)=-i{g\over{2\sqrt{2}}}\sum^{3}_{j,k=1}{U}^{\ast}_{jk}\bar{u}_{k}(x)\bar{\gamma}_{\mu}(1+\gamma_5){d}_{j}(x)
\end{equation}
If $U_{jk}$ is a real parameter with $U^{\ast}_{jk}=U_{jk}$, we get
\begin{equation}
TJ_{\mu+}(x)T^{-1}=J^{+}_{\mu+}(\bar{x})~~~~~~~~~~~~TJ_{\mu-}(x){T}^{-1}=J^{+}_{\mu-}(\bar{x})
\end{equation}
So we have
\begin{equation}
T{\cal{H}}(x)T^{-1}=W^{+}_{\mu}(\bar{x})J^{+}_{\mu+}(\bar{x})+W^{-}_{\mu}(\bar{x})J^{+}_{\mu-}(\bar{x})
\end{equation}
In this cases, we still have $S^{+}_{T}{S}_T=S^{+}{S}$, the transition probability density is unchanged under time reversal. However, as we known that some $U_{jk}$ are complex numbers with $U^{\ast}_{jk}\neq{U}_{jk}$. So in a certain cases, we would have $TJ_{\mu{+}}(x){T}^{-1}\neq{J}^{+}_{\mu{+}}(\bar{x})$ and $TJ_{\mu{-}}(x){T}^{-1}\neq{J}^{+}_{\mu{-}}(\bar{x})$. The result leads to the symmetry violation of time reversal with $S^{+}_T{S}_T\neq{S}^{+}{S}$. The situation is similar to $CP$ violation in weak interaction as shown in Section 5, though $T$ and $CP$ violations have completely same forms are just complementary according to new scheme. At present, $T$ violation has also been founded in some weak interaction processes just as $K^{0}-\bar{K}^{0}$ system's decay and $K_L\rightarrow\pi^{+}\pi^{-}{e}^{+}{e}^{-}$ angle connection $^{(4)}$.
\par
For $QCD$ theory of strong interaction, the interaction Hamiltonian is
\begin{equation}
{\cal{H}}(x)=-B^{\rho}_{\mu}(x)J^{\rho}_{\mu}(x)
\end{equation}
\begin{equation}
J^{\rho}_{\mu}(x)=i{g\over{2}}\sum^{3}_{j,k=1}(\lambda_{\rho})_{jk}\bar{\psi}_j(x)\gamma_{\mu}\psi_{k}(x)
\end{equation}
Here $\bar{\psi}_j=(\bar{\psi}_u,\bar{\psi}_d,\bar{\psi}_s)$ or $(\bar{\psi}_c,\bar{\psi}_b,\bar{\psi}_t)$, $\lambda_{\rho}$ is the Gell-Mann matrix. Because the Gell-mann matrix elements are either real numbers or pure imaginary numbers, according to the current theory, we have $T(\lambda_{\rho})_{jk}{T}^{-1}=(\lambda^{\ast}_{\rho})_{jk}=(\pm\lambda_{\rho})_{jk}$ and get
\begin{equation}
TJ^{\rho}_{\mu}(\vec{x},t)T^{-1}=-i{g\over{2}}\sum(\pm\lambda_{\rho})_{jk}\bar{\psi}_{j}(\vec{x},-t)\gamma_{\mu}\psi_{k}(\vec{x},-t)
\end{equation}
But according to new scheme, $\lambda_{\rho}$ has nothing to do with $T$ transformation. By considering relation $(\lambda_{\rho})_{jk}=(\pm\lambda_{\rho})_{kj}$, we have 
$$TJ^{\rho}_{\mu}(x)T^{-1}=-i{g\over{2}}\sum^{3}_{j,k=1}(\lambda_{\rho})_{jk}\bar{\psi}_k(\bar{x})\bar{\gamma}_{\mu}\psi_j(\bar{x})$$
\begin{equation}
=-i{g\over{2}}\sum^{3}_{j,k=1}(\pm\lambda_{\rho})_{kj}\bar{\psi}_k(\bar{x})\bar{\gamma}_{\mu}\psi_j(\bar{x})=-i{g\over{2}}\sum^{3}_{j,k=1}(\pm\lambda_{\rho})_{jk}\bar{\psi}_j(\bar{x})\bar{\gamma}_{\mu}\psi_{k}(\bar{x})
\end{equation}
Similar to the situation of electromagnetic interaction, it is obvious that when concrete processes are calculated, the results of both schemes are the same. The transition probability densities are invariable under time reversal when regularization and renormalization are not considered in high order processes. 
\\
\\
{\bf 3. P Transformation}
\par
According to the current theory, we have $\vec{x}\rightarrow{-\vec{x}}$ under $P$ transformation. So $P$ transformations of scalar, spinor and electromagnetic fields are defined individually in the current theory
\begin{equation}
P\varphi(\vec{x},t)P^{-1}=-\varphi(-\vec{x},t)~~~~~~~~~P\psi(\vec{x},t)P^{-1}=\gamma_4\psi(-\vec{x},t)
\end{equation}
\begin{equation}
P\vec{A}(\vec{x},t)P^{-1}=-\vec{A}(-\vec{x},t)~~~~~~~~~~PA_4(\vec{x},t)P^{-1}=A_4(-\vec{x},t)
\end{equation}
The Hamiltonian of electromagnetic and strong interactions are unchanged under $P$ transformation with 
\begin{equation}
P{\cal{H}}(\vec{x},t)P^{-1}={\cal{H}}(-\vec{x},t)
\end{equation}
But the Hamiltonian of weak interaction can't keep unchanged under $P$. For quantized scalar fields, $P$ transformation is carried out according to following procedure. From Eqs. (16) and (127), we have
$$P\varphi(\vec{x},t)P^{-1}=-{1\over{(2\pi)^{3/2}}}\int^{+\infty}_{-\infty}{{d^3\vec{p}}\over{\sqrt{2E}}}[a^{+}(\vec{p})e^{-i(-\vec{p}\cdot\vec{x}-Et)}+a(\vec{p})e^{i(-\vec{p}\cdot\vec{x}-Et)}]$$
\begin{equation}
=-{1\over{(2\pi)^{3/2}}}\int^{+\infty}_{-\infty}{{d^3\vec{p}}\over{\sqrt{2E}}}[a^{+}(-\vec{p})e^{-i(\vec{p}\cdot\vec{x}-Et)}+a(-\vec{p})e^{i(\vec{p}\cdot\vec{x}-Et)}]
\end{equation}
As for why $P$ operator only changes the direction of coordinate $\vec{x}$, but not change the directions of momentum $\vec{p}$ in operators $a^{+}(\vec{p})$ and $a(\vec{p})$ as well as function $\exp[i(\vec{p}\cdot\vec{x}-Et)]$ simultaneously, there is no any physical and logical explanation. On the other hand, it is thought that operator $P$ is only acted on creation and annihilation operators, we have
\begin{equation}
P\varphi(\vec{x},t)P^{-1}=-{1\over{(2\pi)^{3/2}}}\int^{+\infty}_{-\infty}{{d^3\vec{p}}\over{\sqrt{2E}}}[Pa^{+}(\vec{p})P^{-1}e^{-i(\vec{p}\cdot\vec{x}-Et)}+Pa(\vec{p})P^{-1}e^{i(\vec{p}\cdot\vec{x}-Et)}]
\end{equation}
Comparing Eqs.(130) and (131), we get $Pa^{+}(\vec{p})P^{-1}=a^{+}(-\vec{p})$, $P\alpha(\vec{p})P^{-1}=\alpha(-\vec{p})$. 
\par
According to new scheme, the basic nature of $P$ transformation is to let $\vec{x}\rightarrow{-\vec{x}}$ and $\vec{p}\rightarrow{-\vec{p}}$, in operators and other relative functions. 
\par
The concrete transformations are discussed below. As shown in the current theory, because $P^{2}\sim{1}$, we have $P\sim\pm{1}$ and take $P\sim-1$. Let $\vec{x}\rightarrow{-\vec{x}}$ and $\vec{p}\rightarrow{-\vec{p}}$,  $P$ transformation of free real scalar field is in light of new scheme
\begin{equation}
P\varphi(\vec{x},t)=-Ae^{i(\vec{p}\cdot\vec{x}-Et)}=-\varphi(\vec{x},t)
\end{equation}
The result is different from the current one shown in Eq.(127). For quantized free scalar field, according to new scheme, by taking $P\sim-1$, we have
\begin{equation}
P\varphi^{(+)}(\vec{x},t)P^{-1}=-{1\over{(2\pi)^{3/2}}}\int^{-\infty}_{+\infty}{{d^3(-\vec{p})}\over{\sqrt{2E}}}{a}^{+}(-\vec{p})e^{-i(\vec{p}\cdot\vec{x}-Et)}
\end{equation}
Let $-\vec{p}\rightarrow{\vec{p}}$  again, the formula above becomes
\begin{equation}
P\varphi^{(+)}(\vec{x},t)P^{-1}=-{1\over{(2\pi)^{3/2}}}\int^{+\infty}_{-\infty}{{d^3\vec{p}}\over{\sqrt{2E}}}{a}^{+}(\vec{p})e^{-i(-\vec{p}\cdot\vec{x}-Et)}=-\varphi^{(+)}(-\vec{x},t)=-\varphi^{(+)}(\bar{x})
\end{equation}
Similarly, we have
\begin{equation}
P\varphi^{(-)}(x)P^{-1}=-\varphi^{(-)}(\bar{x})
\end{equation}
So we have at last
\begin{equation}
P\varphi(x)P^{-1}=P[\varphi^{(+)}(x)+\varphi^{(-)}(x)]P^{-1}=-\varphi(\bar{x})
\end{equation}
It is the same as the current result shown in Eq.(127). For quantized free complex scalar fields, we can get by the same method
\begin{equation}
P\varphi(x)P^{-1}=-\varphi(\bar{x})~~~~~~~~~~~~P\varphi^{+}(x)P^{-1}=-\varphi^{+}(\bar{x})
\end{equation}
It is also the same as the current result. The commutation relation of scalar fields is also unchanged under $P$. It can also be proved by the same method shown in time reversal above that we should take $P\theta(t_1-t_2)P^{-1}=\theta(t_1-t_2)$. So the $P$ transformation of scalar field's propagation function is 
\begin{equation}
P\Delta_F(x_1-x_2)P^{-1}=-{i\over{(2\pi)^4}}\int^{+\infty}_{-\infty}{{d^4{p}}\over{p^2+m^2}}e^{ip\cdot(\bar{x}_1-\bar{x}_2)}=\Delta_F(\bar{x}_1-\bar{x}_2)
\end{equation}
$P$ transformation of electromagnetic fields is discussed below. According to the current theory, $P$ transformations of photon's creation and annihilation operators are defined as
\begin{equation}
Pa^{+}_{\sigma}(\vec{k})P^{-1}=a^{+}_{\sigma}(-\vec{k})~~~~~~~~~~~~Pa_{\sigma}(\vec{k})P^{-1}=a_{\sigma}(-\vec{k})
\end{equation}
For free electromagnetic fields, $P$ transformations of polarization vector $\varepsilon^{\sigma}_{\mu}$ are
\begin{equation}
P\epsilon^1_{i}(\vec{k}){P}^{-1}=-\epsilon^1_{i}(\vec{-k})~~~~~P\epsilon^2_{i}(\vec{k}){P}^{-1}=-\epsilon^2_{i}(\vec{-k})~~~~~~P\epsilon^3_{i}(\vec{k}){P}^{-1}=-\epsilon^3_{i}(\vec{-k})~~~~~P\epsilon^4_{4}(\vec{k}){P}^{-1}=\epsilon^4_{4}(\vec{-k})
\end{equation}
The $P$ transformations of electromagnetic potentials are
\begin{equation}
P\vec{A}(x)P^{-1}=-\vec{A}(\bar{x})~~~~~~~~~~~~~~~~PA_4(x)P^{-1}=A_4(\bar{x})
\end{equation}
\par
This kind of $P$ transformations for gauge fields is used commonly in the current theory, but it is worthy of being questioned. According to Eq.(141), $\vec{A}$ is regarded as a vector relative to electric current, so $\vec{A}$ changes its sign under $P$. But $A_4$ is considered as a scalar relative to charge density, it dose not change sign under $P$. However, in the quantum theory of field, quantized electromagnetic field is used to describer photon, in which $A_1$ and $A_2$ are considered to describe transverse photons, while $A_3$ and $A_4$ are considered to describe longitudinal and scalar photons. According to Eq.(141), the parity of transverse photons and longitudinal photons is $-1$ but the parity of scalar photons should be $+1$. However, because photons do not carry charges, it is unnecessary for us to think that $A_4$ describe charge density. We have no any reason to think that the photons described by $A_4$ have something different from the photons described by $\vec{A}$. For the identification of theory, all photon's parity should be the same with value $-1$. At least, longitudinal photon and time should have same parity. Otherwise theory is inconsistent.
\par
To explore this problem more clearly, let's examine classical electromagnetic theory. In the theory, electromagnetic potential $A_{\mu}=(\vec{A},i\varphi)$ satisfies following motion equation
\begin{equation}
\nabla^2{A}_{\mu}-{{\partial^2}\over{\partial{t^2}}}{A}_{\mu}=-4\pi{j}_{\mu}
\end{equation}
Where $j_{\mu}=(\vec{j},i\rho)$. When $j_{\mu}=0$, equation becomes£º
\begin{equation}
\nabla^2{A}_{\mu}-{{\partial^2}\over{\partial{t^2}}}{A}_{\mu}=0
\end{equation}
The general solution of Eq.(142) is
\begin{equation}
A_{\mu}(\vec{x},t)=A^a_{\mu}(\vec{x},t)+A^b_{\mu}(\vec{x},t)
\end{equation}
Where $A^a_{\mu}$ is a special solution satisfying Eq.(142) with form
\begin{equation}
A^a_{\mu}(\vec{x},t)=\int{d}\tau'{{j(\vec{x}',t-r)}\over{r}}
\end{equation}
$A^{b}_{\mu}$ is a common solution satisfying Eq.(143) and the Lorentz condition $\partial_{\mu}{A}^{b}_{\mu}=0$ with form
\begin{equation}
A^b_{\mu}(\vec{x},t)=\int{d}^3\vec{k}[a(\vec{k})e^{ik\cdot{x}}+b(\vec{k})e^{-ik\cdot{x}}]
\end{equation}
It is obvious that Eq.(145) depends on electric current and charge densities, but Eq.(146) has nothing to do with electric current and charges. When we discuss free photon field, there is no charge distribution in space. The photon fields is determined by Eq.(143) and (146), so its $P$ transformation is not restricted by relation $P_{\rho}{P}^{-1}=\rho$. Because $P^2\sim{1}$, $P\sim\pm{1}$, we have $P\epsilon^{\sigma}_4(\vec{k})P^{-1}=\pm\epsilon^{\sigma}_4(-\vec{k})$. So for the identification of theory, we should take $P\epsilon^{\sigma}_4(\vec{k}){P}^{-1}=-\epsilon^{\sigma}_4(-\vec{k})$ or $PA^{b}_4(\vec{x},t)P^{-1}=-A^{b}_4(-\vec{x},t)$. In this way, $P$ transformation of photon fields should be written as
\begin{equation}
PA^{b}_{\mu}(x)P^{-1}=-A^{b}_{\mu}(\bar{x})
\end{equation}
In fact, from the definition of momentum $\vec{p}=d\vec{x}/{dt}$, both $T$ and $P$ transformation lead to $\vec{p}\rightarrow{-\vec{p}}$. It can be said that both $T$ and $P$ transformation are equivalent from the angle of momentum space. We always regard the processes of time reversal as one in which the directions of all particle's velocities are reversed. But the $P$ transformation of fields $A^{a}_{\mu}(x)$ should still be described by Eq.(141). In the quantum theory of field, electric current density is actually described by spinor fields with form  $j_{\mu}=e\bar{\psi}\gamma_{\mu}\psi$. By considering below relations Eqs.(159) and (164) in new scheme, we have
\begin{equation}
P\vec{j}(x)P^{-1}=eP\bar{\psi}(x)P^{-1}{P}\vec{\gamma}{P}^{-1}{P}\psi(x){P}^{-1}=-e\bar{\psi}(\bar{x})\vec{\gamma}\psi(\bar{x})=-\vec{j}(\bar{x})
\end{equation}
\begin{equation}
Pj_4(x)P^{-1}=eP\bar{\psi}(x)P^{-1}{P}\gamma_4{P}^{-1}{P}\psi(x){P}^{-1}=e\bar{\psi}(\bar{x})\gamma_4\psi(\bar{x})=j_4(\bar{x})
\end{equation}
Put them into Eq.(145), we get $PA^{a}_{\mu}(x)P^{-1}=(-\vec{A}^{a}(\bar{x}),A^{a}_4(\bar{x}))=\bar{A}^{a}_{\mu}(\bar{x})$, so we have at last
\begin{equation}
PA_{\mu}(x)P^{-1}=\bar{A}^{a}_{\mu}(\bar{x})-A^{b}_{\mu}(\bar{x})
\end{equation}
\par
As for non-Abelian gauge fields such as $W^{\pm}_{\mu}$, $Z^{0}_{\mu}$ and $B^{\sigma}_{\mu}$, their motion equations are very complex. We have no simple solutions such as Eq.(145) and (146). Because parity is regarded as an inherent nature of micro-particle, for the identification of theory, we should also define $P$ transformation of non-Abelian gauge fields as
\begin{equation}
PW^{\pm}_{\mu}(x)P^{-1}=-W^{\pm}_{\mu}(\bar{x})~~~~~~PZ^{0}_{\mu}(x)P^{-1}=-Z^{0}_{\mu}(\bar{x})~~~~~~~PB^{\rho}_{\mu}(x)P^{-1}=-B^{\rho}_{\mu}(\bar{x})
\end{equation}
\par
The commutation relation of electromagnetic fields is also unchanged under $P$. By the same method, the $P$ transformation of the propagation function of electromagnetic fields (photons) is
\begin{equation}
PD_F(x_1-x_2)_{\mu\nu}{P}^{-1}=-{i\over{(2\pi)^4}}\int^{+\infty}_{-\infty}{d}^4{k}{{\delta_{\mu\nu}}\over{k^2+m^2}}{e}^{ik\cdot(\bar{x}_1-\bar{x}_2)}=D_F(\bar{x}_1-\bar{x}_2)_{\mu\nu}
\end{equation}
\par
For quantized spinor fields, the spins of particles are unchanged under $P$, but their helicities would change with $s\rightarrow-s$. In new scheme, we also define $P$ transformations of sipinor fields and their creation and annihilation operators as
\begin{equation}
P\bar{u}_s(\vec{p})P^{-1}=\bar{u}_{-s}(-\vec{p})~~~~~~~~~~~~~~~~~~Pu_s(\vec{p})P^{-1}=u_{-s}(-\vec{p})
\end{equation}
\begin{equation}
P\bar{\nu}_s(\vec{p})P^{-1}=\bar{\nu}_{-s}(-\vec{p})~~~~~~~~~~~~~~~~~~P\nu_s(\vec{p})P^{-1}=\nu_{-s}(-\vec{p})
\end{equation}
\begin{equation}
Pb^{(+)}_s(\vec{p})P^{-1}=b^{(+)}_{-s}(-\vec{p})~~~~~~~~~~~~~~~~~~Pb_s(\vec{p})P^{-1}=b_{-s}(-\vec{p})
\end{equation}
\begin{equation}
Pd^{(+)}_s(\vec{p})P^{-1}=d^{(+)}_{-s}(-\vec{p})~~~~~~~~~~~~~~~~~~Pd_s(\vec{p})P^{-1}=d_{-s}(-\vec{p})
\end{equation}
But from the definitions above and by the same method, we can get
\begin{equation}
P\bar{\psi}^{(+)}_s(x)P^{-1}=\bar{\psi}^{(+)}_{-s}(\bar{x})~~~~~~~~~~~~~~~~~~P\psi^{(-)}_s(x)P^{-1}=\psi^{(-)}_{-s}(\bar{x})
\end{equation}
\begin{equation}
P\bar{\psi}^{(-)}_s(x)P^{-1}=\bar{\psi}^{(-)}_{-s}(\bar{x})~~~~~~~~~~~~~~~~~~P\psi^{(+)}_s(x)P^{-1}=\psi^{(+)}_{-s}(\bar{x})
\end{equation}
Because it is the same to take sum over $s$ and $-s$, we get at last
\begin{equation}
P\bar{\psi}(x)P^{-1}=\sum_{s}\bar{\psi}_{-s}(\bar{x})=\bar{\psi}(\bar{x})~~~~~~~~~~~~~~~~~~P\psi(x)P^{-1}=\sum_{s}\psi_{-s}(\bar{x})=\psi(\bar{x})
\end{equation}
The result is quite different from Eq.(127). The commutation relation of spinor fields is also unchanged under $P$, so $P$ transformation of propagation function of spinor fields is unchanged with
\begin{equation}
PS_F(x_1-x_2)_{\alpha\beta}{P}^{-1}={i\over{(2\pi)^4}}\int^{+\infty}_{-\infty}{d}^4{p}({{m-i\hat{p}}\over{p^2+m^2}})_{\alpha\beta}{e}^{ip\cdot(\bar{x}_1-\bar{x}_2)}=S_F(\bar{x}_1-\bar{x}_2)_{\alpha\beta}
\end{equation}
\par
Now let's discuss the matrix form of operator $P$ in new scheme. By acting operator $P$ on the motion equation of positive particle in momentum space, and by considering Eq.(153) and relations $P\vec{p}{P}^{-1}=-\vec{p}$, $Pp_4{P}^{-1}=p_4$, we can get
\begin{equation}
P(i\hat{p}+m){u}_s(\vec{p})P^{-1}=(-iP\vec{\gamma}{P}^{-1}\cdot\vec{p}+iP\gamma_{4}{P}^{-1}{p}_4)u_{-s}(-\vec{p})=0
\end{equation}
On the other hand, as discussed in time reversal, we have relations $u_{-s}(-\vec{p})=u_{-s}(\vec{p})$, $\nu_{-s}(-\vec{p})=\nu_{-s}(\vec{p})$. 
Thus, Eq.(161) can be written as
\begin{equation}
(-iP\vec{\gamma}{P}^{-1}\cdot\vec{p}+iP\gamma_4{P}^{-1}p_4)u_{-s}(\vec{p})=0
\end{equation}
Because $u_{-s}(\vec{p})$ and $u_s(\vec{p})$ satisfy the same motion equation $(i\vec{\gamma}\cdot\vec{p}+i\gamma_4{p}_4)u_{\pm{s}}(\vec{p})=0$ , we have
\begin{equation}
P\vec{\gamma}P^{-1}=-\vec{\gamma}~~~~~~~~~P\gamma_4{P}^{-1}=\gamma_4~~~~~~~~or~~~~~~~~P\gamma_{\mu}{P}^{-1}=\bar{\gamma}_{\mu}
\end{equation}
Also as the current theory, we can take
\begin{equation}
P=\pm\gamma_4
\end{equation}
\par
Now let us discuss the $P$ transformation of the Hamiltonian of electromagnetic interaction. By means of Eq.(147), (159) and (164), we have
\begin{equation}
P{\cal{H}}(\vec{x},t)(P)^{-1}=-{{ie}\over{2}}A_{\mu}(x)[\bar{\psi}(\bar{x})\bar{\gamma}_{\mu}\psi(\bar{x})-\psi^{\tau}(\bar{x})\bar{\gamma}^{\tau}_{\mu}\bar{\psi}(\bar{x})^{\tau}]
\end{equation}
Comparing with Eq.(129), the difference is at $\bar{\gamma}_{\mu}\rightarrow\gamma_{\mu}$ besides there is a negative sign less. The form of $P$ transformation is the same as $T$ transformation shown in Eq.(75) besides a difference of a negative sign. But it should note that there exists process's reversion in $T$ transformation and there is no process's reversion in $P$ transformation. It is also easy to prove that the transformation shown in Eq.(165) does not change transition probability densities.
\par
At first, for the processes with photon propagation lines in Feynman diagrams, for example, electron-electron scatting, transition probability amplitude is shown in Eq.(78). By means of relation $\bar{\gamma}_{\mu}\bar{\gamma}_{\mu}=\gamma_{\mu}\gamma_{\mu}$, it is easy to prove $S_P=PSP^{-1}=-S$ and $S^{+}_P{S}_P=S^{+}{S}$ under the transformation of Eq.(172). So transition probability density is invariable. For the processes with electron propagation lines in Feynman diagrams, for example, the Compton scattering described by Eq.(81), we have $S_P\neq\pm{S}$ under the transformation of Eq.(165). But statistical average about photon's polarizations should be considered when transition probability densities are calculated. By using relation $\bar{\gamma}_{\mu}\bar{\gamma}_{\mu}=\gamma_{\mu}\gamma_{\mu}$  and
\begin{equation}
\sum^{4}_{\sigma=1}\epsilon^{\sigma}_{\mu}\epsilon^{\sigma}_{\nu}=\delta_{\mu\nu}
\end{equation}
we can get $S^{+}_{P}{S}_{P}=S^{+}{S}$. So transition probability density is still unchanged. The conclusion is suitable for high order processes.
\par
   For $QCD$ theory of strong interaction, because matrix $\lambda_{\rho}$ does not appear in the equation of free quark field, we can consider that $\lambda_{\rho}$ has nothing to do with $P$ transformation with $P\lambda_{\rho}{P}^{-1}=\lambda_{\rho}$. By means of Eqs.(151), (159) and (164), we get
\begin{equation}
P{\cal{H}}(x){P}^{-1}=i{g\over{2}}\sum_{j,k}(\lambda_{\rho})_{jk}\bar{\psi}_j(\bar{x})\bar{\gamma}_{\mu}{B}^{\rho}_{\mu}(\bar{x})\psi_k(\bar{x})
\end{equation}
The difference is at $\bar{\gamma}_{\mu}=\gamma_{\mu}$ besides there is a negative sign less comparing with the current theory. The situation is similar to electromagnetic interaction with photon propagation lines in the processes. By relation $\bar{\gamma}_{\mu}\bar{\gamma}_{\mu}=\gamma_{\mu}\gamma_{\mu}$, we can also prove the invariability of transition probability densities under the transformation of Eq.(165). 
\par
For weak interaction between leptons, by considering Eq.(159) and (164), we have
\begin{equation}
P\gamma_5{P}^{-1}=-\gamma_5
\end{equation}
\begin{equation}
P\bar{\psi}_{\nu}(x)\gamma_{\mu}(1+\gamma_5)\psi_{l}(x){P}^{-1}=\bar{\psi}_{\nu}(\bar{x})\bar{\gamma}_{\mu}(1-\gamma_5)\psi_{l}(\bar{x})
\end{equation}
So we get
\begin{equation}
P{\cal{H}}_w(x){P}^{-1}=i{g\over{2\sqrt{2}}}[W^{+}_{\mu}(\bar{x})\bar{\psi}_{\nu}(\bar{x})\bar{\gamma}_{\mu}(1-\gamma_5)\psi_l(\bar{x})+W^{-}_{\mu}(\bar{x})\bar{\psi}_l(\bar{x})\bar{\gamma}_{\mu}(1-\gamma_5)\psi_{\nu}(\bar{x})]
\end{equation}
\begin{equation}
P{\cal{H}}_z(x){P}^{-1}=i{{\sqrt{g^2+g'^2}}\over{4}}{Z}^{0}_{\mu}[\bar{\psi}_{\nu}(\bar{x})\bar{\gamma}_{\mu}(1-\gamma_5)\psi_{\nu}(\bar{x})+\bar{\psi}(\bar{x})\bar{\gamma}_{\mu}(4\sin^2\theta_W-1-\gamma_5)\psi_l(\bar{x})]
\end{equation}
The difference is also at $\bar{\gamma}_{\mu}\rightarrow\gamma_{\mu}$ besides there is a negative sign less. By using relation $\bar{\gamma}_{\mu}\bar{\gamma}_{\mu}=\gamma_{\mu}\gamma_{\mu}$, we know that the transition probability density is same as that in the current theory with parity symmetry violation. For the weak interaction between quarks, we get
\begin{equation}
PJ_{\mu+}(x){P}^{-1}=i{g\over{2\sqrt{2}}}\sum^{N}_{j,k=1}{U}_{jk}\bar{u}_j(\bar{x})\bar{\gamma}_{\mu}(1-\gamma_5){d}_k(\bar{x})
\end{equation}
\begin{equation}
PJ_{\mu-}(x){P}^{-1}=i{g\over{2\sqrt{2}}}\sum^{N}_{j,k=1}{U}_{jk}\bar{d}_j(\bar{x})\bar{\gamma}_{\mu}(1-\gamma_5){u}_k(\bar{x})
\end{equation}
The difference is also at $\bar{\gamma}_{\mu}\rightarrow\gamma_{\mu}$, as well as a negative sign less. When transition probabilities are calculated, there exists the same parity symmetry violation as appearing in the current theory.
\\
\\
{\bf 4. C Transformation}
\par
According to the current theory, the $C$ transformations of complex scalar fields and electromagnetic field are defined as follows
\begin{equation}
CA_{\mu}(x)C^{-1}=-A_{\mu}(x)~~~~~~~~C\varphi^{+}(x)C^{-1}=\varphi(x)~~~~~~~~C\varphi(x)C^{-1}=\varphi^{+}(x)
\end{equation}
The motion equation of a spinor positive particle with charge $q$ in electromagnetic field $A_{\mu}(x)$ is
\begin{equation}
[r_{\mu}\partial_{\mu}-iq\gamma_{\mu}{A}_{\mu}(x)+m]\psi(x)=0
\end{equation}
The motion equation of a spinor anti-particle with charge $-e$ moving in the same field is
\begin{equation}
[r_{\mu}\partial_{\mu}+iq\gamma_{\mu}{A}_{\mu}(x)+m]\psi_c(x)=0
\end{equation}
When operator $C$ satisfies relation
\begin{equation}
C\gamma^{\tau}_{\mu}{C}^{-1}=-\gamma_{\mu}~~~~~~or~~~~~C\gamma_{\mu}{C}^{-1}=-\gamma^{\tau}_{\mu}=(\gamma_1,-\gamma_2,\gamma_3,-\gamma_4)
\end{equation}
The wave functions of spinor positive and ant-anti-particles satisfy following transformation
\begin{equation}
\psi_c=C\bar{\psi}^{\tau}=\gamma_2\gamma_4\bar{\psi}^{\tau}=\gamma_2\psi^{\ast}~~~~~~~~~~~~~\bar{\psi}_c=\psi^{\tau}\gamma_2\gamma_4
\end{equation}
Eq.(176) can be chansformed into Eq.(175). However, in the current documents, Eq.(178) always be written as the partial form with
\begin{equation}
(\psi_c)_{\alpha}=(\gamma_2)_{\alpha\beta}\psi^{+}_{\beta}~~~~~~~~~~~(\psi^{+}_c)_{\alpha}=\psi_{\beta}(\gamma_2)_{\beta\alpha}
\end{equation}
This is improper for the formulas actually means $\psi_c=\gamma_2\psi^{+}$, and $\psi^{+}_c=\psi\gamma_2$. The product rule of matrix is not satisfied, besides Eq.(179) has different meaning comparing with Eq.(178). 
\par
The more serious problem is that in the quantum theory of field, we define $\psi=\psi^{(+)}+\psi^{(-)}$ and $\bar{\psi}^{(+)}=\bar{\psi}^{(+)}+\bar{\psi}^{(-)}$. The second quantization operators $\psi$ and $\bar{\psi}$ have contained both the positive particle's components $\psi^{(-)}$ and $\bar{\psi}^{(+)}$ as well as the anti-particle's components $\psi^{(+)}$ and $\bar{\psi}^{(-)}$. It is improper to regard $\psi$ and $\bar{\psi}$ as the wave function of positive particles again. Similarly, after $C$ transformation, $\psi_c$ and $\bar{\psi}_c$ have also contained the both components of positive particle and anti-particle. It is also improper to regard $\psi_c$ and  $\bar{\psi}_c$  as the wave function of positive particles again. The real meaning of $C$ transformation in the quantum theory of field should be that in the coordinate space the creation operator $\bar{\psi}^{(+)}_s(x)$ of spinor positive particle exchange with the creation operator $\psi^{(+)}_s(x)$ of spinor anti-particle. The annihilation operator $\psi^{(-)}_s(x)$ of spinor positive particle exchanges with the annihilation operator $\bar{\psi}^{(-)}_s(x)$ of spinor anti-particle. 
\par
By the consideration above, we can establish a new $C$ transformation theory based on following fundamental natures of C transformation.
\par
1. The creation operators of positive particle and anti-particle exchange each other, and the annihilation operators of positive particle and anti-particle also exchange each other.
\par
2. The wave function of spinor positive particle in momentum space is transformed into conjugate wave function of spinor anti-particles. The wave function of spinor anti-particle in momentum space is transformed into conjugate wave function of spinor positive particles:
\begin{equation}
Cu_{s\alpha}(\vec{p})C^{-1}=\bar{\nu}_{s\alpha}(\vec{p})~~~~~~~~~~or~~~~~~~~~Cu_s(\vec{p})C^{-1}=\bar{\nu}^{\tau}_{s}(\vec{p})
\end{equation}
\begin{equation}
C\nu_{s\alpha}(\vec{p})C^{-1}=\bar{u}_{s\alpha}(\vec{p})~~~~~~~~~~or~~~~~~~~~C\nu_s(\vec{p})C^{-1}=\bar{u}^{\tau}_{s}(\vec{p})
\end{equation}
\par
For free real scalar field, the wave function of anti-particle is itself and is unchanged under$C$. For quantized free complex scalar field, $b^{+}(\vec{p})$ is the creation operator of anti-particle and $a^{+}(\vec{p})$ is the creation operator of positive particle. We have $Cb^{+}(\vec{p}){C}^{-1}=a^{+}(\vec{p})$, so
$$C\varphi^{(+)}(x)C^{-1}={1\over{(2\pi)^{3/2}}}\int^{+\infty}_{-\infty}{{d^3\vec{p}}\over{\sqrt{2E}}}{C}b^{+}(\vec{p}){C}^{-1}{e}^{-ip\cdot{x}}$$
\begin{equation}
={1\over{(2\pi)^{3/2}}}\int^{+\infty}_{-\infty}{{d^3\vec{p}}\over{\sqrt{2E}}}a^{+}(\vec{p}){e}^{-ip\cdot{x}}=\varphi^{+(+)}(x)
\end{equation}
Similarly we have $Cb(\vec{p})C^{-1}=a(\vec{p})$ and 
\begin{equation}
C\varphi^{(-)}(x)C^{-1}=\varphi^{+(-)}(x)
\end{equation}
Thus we get $C\varphi(x)C^{-1}=\varphi^{+}(x)$. The result is the same as the current theory shown in Eq.(174). 
\par
The propagation function of scalar field is discussed below. By considering Eq.(29) and results above,  we get
\begin{equation}
C\Delta_F(x_1-x_2)C^{-1}=C\theta(t_1-t_2)C^{-1}[\varphi^{+(-)}(x_1),\varphi^{(+)}(x_2)]+C\theta(t_2-t_1)C^{-1}[\varphi^{(-)}(x_2),\varphi^{+(+)}(x_1)]
\end{equation}
Because $C^2\sim{1}$, $C\sim\pm{1}$£¬ we have $C\theta(t_1-t_2)C^{-1}=\pm\theta(t_1-t_2)$. Taking $C\theta(t_1-t_2)C^{-1}=\theta(t_1-t_2)$, we get
\begin{equation}
C\Delta_F(x_1-x_2)C^{-1}=\theta(t_1-t_2)[\varphi^{+(-)}(x_1),\varphi^{(+)}(x_2)]+\theta(t_2-t_1)[\varphi^{(-)}(x_2),\varphi^{(+)}(x_1)]
\end{equation}
It is equivalent to let $x_1\leftrightarrow{x}_2$ on the right side of Eq.(29), so the $C$ transformation of scalar field's propagation function is
\begin{equation}
C\Delta_F(x_1-x_2)C^{-1}=-{i\over{(2\pi)^4}}\int^{+\infty}_{-\infty}{{d^4{k}}\over{k^2+m^2}}{e}^{ik\cdot(x_2-x_1)}=\Delta_F(x_2-x_1)
\end{equation}
The $C$ transformation of electromagnetic field is the same as the current theory shown in Eq.(174). The $C$ transformation of electromagnetic field's propagation function is also similar to Eq.(186). For the quantized spinor fields, the motion equations and their conjugate equations of positive and anti-particles in both coordinate and momentum spaces are individually 
\begin{equation}
(\gamma_{\mu}\partial_{\mu}+m)\psi^{(-)}_s(x)=0~~~~~~~~~~~~(i\hat{p}+m)u_s(\vec{p})=0
\end{equation}
\begin{equation}
(\gamma_{\mu}\partial_{\mu}+m)\psi^{(+)}_s(x)=0~~~~~~~~~~~~(-i\hat{p}+m)\nu_s(\vec{p})=0
\end{equation}
\begin{equation}
\bar{\psi}^{(+)}_s(x)(\gamma_{\mu}\partial_{\mu}-m)=0~~~~~~~~~~~~\bar{u}_s(\vec{p})(i\hat{p}+m)=0
\end{equation}
\begin{equation}
\bar{\psi}^{(-)}_s(x)(\gamma_{\mu}\partial_{\mu}-m)=0~~~~~~~~~~~~\bar{\nu}_s(\vec{p})(-i\hat{p}+m)=0
\end{equation}
According to new scheme, we have 
$$C\psi^{(+)}_{s\alpha}(x)C^{-1}={1\over{(2\pi)^{3/2}}}\int^{+\infty}_{-\infty}{d}^3\vec{p}\sqrt{{m\over{E}}}{C}\nu_{s\alpha}(\vec{p})C^{-1}{C}d^{+}_s(\vec{p})C^{-1}{e}^{-ipx}$$
\begin{equation}
={1\over{(2\pi)^{3/2}}}\int^{+\infty}_{-\infty}{d}^3\vec{p}\sqrt{{m\over{E}}}\bar{u}^{\tau}_{s\alpha}(p)b^{+}_s(\vec{p}){e}^{-ipx}=\bar{\psi}^{(+)\tau}_{s\alpha}(x)
\end{equation}
or $C\psi^{(+)}_s(x){C}^{-1}=\bar{\psi}^{(+)\tau}_s(x)$. Be all the same, we have $C\bar{\psi}^{(+)}_s(x){C}^{-1}=\psi^{(+)\tau}_s(x)$, as well as
\begin{equation}
C\psi^{(-)}_{s }(x)C^{-1}=\bar{\psi}^{(-)\tau}_{s}(x)~~~~~~~~~~~~C\bar{\psi}^{(-)}_s(x)C^{-1}=\psi^{(-)\tau}_s(x)
\end{equation}
At last, we get
\begin{equation}
C\bar{\psi}(x)C^{-1}=\psi^{\tau}(x)~~~~~~~~~~~~C\psi(x)C^{-1}=\bar{\psi}^{\tau}(x)
\end{equation}
It means that $\bar{\psi}(x)$ and $\psi^{\tau}(x)$ exchange to each other under $C$. The result is different from the current $C$ transformation shown in Eq.(178), but similar to new time reversal transformation.
\par
   The matrix form of operator $C$ is discussed below. The $C$ transformation of motion equation of spinor anti-particle in momentum space is
\begin{equation}
(-iC\gamma_{\mu}{C}^{-1}p_{\mu}+m)C\nu_s(\vec{p}){C}^{-1}=0
\end{equation}
By using Eq.(181), we get
\begin{equation}
\bar{u}_s(\vec{p})[-i(C\gamma_{\mu}{C}^{-1})^{\tau}{p}_{\mu}+m]=0
\end{equation}
Comparing it with Eq.(189), we get $(Cr_{\mu}{C}^{-1})^{\tau}=-\gamma_{\mu}$. The result is the same as the current theory shown in Eq.(177). We can take
\begin{equation}
C=\pm{i}\gamma_2\gamma_4
\end{equation}
\par
The $C$ transformation of the propagation function of spinor field is discussed below. As shown before, we should take $C\theta(t_1-t_2){C}^{-1}=\theta(t_1-t_2)$. From Eq.(191) and (192), we have
\begin{equation}
CS_F(x_1-x_2)_{\alpha\beta}{C}^{-1}=\theta(t_1-t_2)\{\bar{\psi}^{(-)}_{\alpha}(x_1),\psi^{(+)}_{\beta}(\bar{x})\}-\theta(t_2-t_1)\{\psi^{(-)}_{\beta}(x_2),\bar{\psi}^{(+)}_{\alpha}(x_1)\}
\end{equation}
It is equal to let $x_1\leftrightarrow{x}_2$, $\alpha\leftrightarrow\beta$ on the right side of Eq.(59) besides a negative sign is added, so we have
\begin{equation}
CS_F(x_1-x_2)_{\alpha\beta}{C}^{-1}={i\over{(2\pi)^4}}\int^{+\infty}_{-\infty}{d}^4{p}({{m-i\hat{p}}\over{p^2+m^2}})_{\beta\alpha}{e}^{ip\cdot(x_2-x_1)}=-S_F(x_2-x_1)_{\beta\alpha}
\end{equation}
\par
So the propagation function of spinor field would change a negative sign under $C$, similar to Eq.(62) in new $T$ transformation.Thus, the $C$ transformation of electromagnetic interaction is 
$$C{\cal{H}}(x)C^{-1}=i{e\over{2}}A_{\mu}(x)[\psi^{\tau}(x)(-\gamma^{\tau}_{\mu})\bar{\psi}^{\tau}(x)-\bar{\psi}(x)(-\gamma_{\mu})\psi(x)]$$
\begin{equation}
=i{e\over{2}}A_{\mu}(x)[\bar{\psi}(x)\gamma_{\mu}\psi(x)-\psi^{\tau}(x)\gamma^{\tau}_{\mu}\bar{\psi}^{\tau}(x)]
\end{equation}
There is a difference of negative sign comparing with the result of current transformation theory, but transition probability density is unchanged. It is known that according to new scheme, the differences between $T$ and $C$ transformations are at $\bar{\gamma}_{\mu}\rightarrow-\gamma_{\mu}$ (or $\gamma_4\rightarrow-\gamma_4$) and $\bar{x}\rightarrow{x}$. But this kind of difference does not affect transition probability. 
\par
For the united theory of electro-weak interaction between leptons, we have $C\gamma_5{C}^{-1}=\gamma_5$ from Eq.(196). By the results shown in Eqs.(177) and (193), we have
$$C\bar{\psi}_{\nu}\gamma_{\mu}(1+\gamma_5)\psi_l{C}^{-1}=C\bar{\psi}_{\nu}(1-\gamma_5)\gamma_{\mu}\psi_l{C}^{-1}=\psi_{\nu\alpha}(1-\gamma_5)_{\alpha\beta}(-\gamma^{\tau}_{\mu})_{\beta\sigma}\bar{\psi}_{l\sigma}$$
\begin{equation}
=\bar{\psi}_{l\sigma}(\gamma_{\mu})_{\sigma\beta}(1-\gamma^{\tau}_5)_{\beta\alpha}\psi_{\nu\alpha}=\bar{\psi}_l{r}_{\mu}(1-\gamma_5)\psi_{\nu}
\end{equation}
Similarly, we get
\begin{equation}
C\bar{\psi}_l\gamma_{\mu}(1+\gamma_5)\psi_{\nu}{C}^{-1}=\bar{\psi}_{\nu}\gamma_{\mu}(1-\gamma_5)\psi_l
\end{equation}
At the same time, we also have
\begin{equation}
CW^{\pm}_{\mu}(x){C}^{-1}=-W^{\mp}_{\mu}(x)~~~~~~~CZ^{0}_{\mu}(x)T^{-1}=-Z^{0}_{\mu}(x)
\end{equation}
The $C$ transformations of the Hamiltonians of weak interactions between leptons are
\begin{equation}
C{\cal{H}}_w(x)C^{-1}=i{g\over{2\sqrt{2}}}[W^{+}_{\mu}(x)\bar{\psi}_{\nu}(x)\gamma_{\mu}(1-\gamma_5)\psi_l(x)+W^{-}_{\mu}(x)\bar{\psi}_l(x)\gamma_{\mu}(1-\gamma_5)\psi_{\nu}(x)]
\end{equation}
\begin{equation}
C{\cal{H}}_z(x)C^{-1}=i{{\sqrt{g^2+g'^2}}\over{4}}{Z}^0_{\mu}(x)[\bar{\psi}_{\nu}(x)\gamma_{\mu}(1-\gamma_5)\psi_{\nu}(x)+\bar{\psi}_l(x)\gamma_{\mu}(4{\sin}^2\theta_W-1-\gamma_5)\psi_l(x)]
\end{equation}
There exists a difference of negative sign comparing with the current theory. The Hamiltonians of weak interactions between leptons can't keep unchanged under $C$. The situation is similar for the weak interaction between quarks.
\par
The $C$ transformation of the $QCD$ theory of strong interaction is discussed at last. Because $C$ transformation has nothing to do with $\lambda_{\rho}$, we have $C\lambda_{\rho}{C}^{-1}=\lambda_{\rho}$. Considering the fact that the Gell-mann matrix elements have the nature $(\lambda_{\rho})_{jk}=\pm(\lambda_{\rho})_{kj}$, by the relation $CB^{\rho}_{\mu}{C}^{-1}=-B^{\rho}_{\mu}$ and Eq.(177), we have
$$C{\cal{H}}(x)C^{-1}=i{g\over{2}}\sum^3_{j,k=1}(\lambda_{\rho})_{jk}\psi_{j\alpha}(x)(-\gamma^{\tau}_{\mu})_{\alpha\beta}{B}^{\rho}_{\mu}(x)\bar{\psi}_{k\beta}(x)$$
\begin{equation}
=i{g\over{2}}\sum^3_{j,k=1}(\lambda_{\rho})_{jk}\bar{\psi}_k(x)\gamma_{\mu}{B}^{\rho}_{\mu}(x)\psi_j(x)=i{g\over{2}}\sum^3_{j,k=1}(\pm\lambda_{\rho})_{kj}\bar{\psi}_k(x)\gamma_{\mu}{B}^{\rho}_{\mu}(x)\psi_j(x)
\end{equation}
According to the current theory, we have
\begin{equation}
\psi_c=C\bar{\psi}^{\tau}~~~~~~~~\bar{\psi}_c=-\psi^{\tau}{C}^{-1}~~~~~~~~B^{\rho}_{\mu{c}}=-B^{\rho}_{\mu}~~~~~~C^{-1}\gamma^{\tau}_{\mu}{C}=-\gamma_{\mu}
\end{equation}
$${\cal{H}}_c(x)=-i{g\over{2}}\sum^3_{j,k=1}(\lambda_{\rho})_{jk}\bar{\psi}_{jc}(x)\gamma_{\mu}{B}^{\rho}_{\mu{c}}(x)\psi_{kc}(x)$$
\begin{equation}
=-i{g\over{2}}\sum^3_{j,k=1}(\lambda_{\rho})_{jk}\bar{\psi}_k(x)\gamma_{\mu}{B}^{\rho}_{\mu}(x)\psi_j(x)=-i{g\over{2}}\sum^3_{j,k=1}(\pm\lambda_{\rho})_{kj}\bar{\psi}_k(x)\gamma_{\mu}{B}^{\rho}_{\mu}(x)\psi_j(x)
\end{equation}
There is only a difference of negative sign comparing with Eq.(205), so the transition probability densities are the same. Similarly, according to new scheme, if regularization and renormalization in high order processes are considered, the symmetry of $C$ transformation would be violated. This problem will be discussed in Section 6.
\\
\\
{\bf 5. United Transformations}
\par 
Therefore, in the new scheme, we have the united $CPT$ transformations
\begin{equation}
CPT=(\pm{i}\gamma_2\gamma_4)(\pm\gamma_4)(\pm{i}\gamma_1\gamma_3\gamma_4)=\pm\gamma_5~~~~~~~~~ CPT\gamma_{\mu}(CPT)^{-1}=-\gamma_{\mu}
\end{equation}
\begin{equation}
CPT\varphi(x)(CPT)^{-1}=-\varphi(x)~~~~~~~~~~~~~CPT\varphi^{+}(x)(CPT)^{-1}=-\varphi^{+}(x)
\end{equation}
\begin{equation}
CPTA_{\mu}(x)(CPT)^{-1}=-A_{\mu}(x)~~~~~~~~~~~~CPTB^{\rho}_{\mu}(x)(CPT)^{-1}=-B^{\rho}_{\mu}(x)
\end{equation}
\begin{equation}
CPTW^{\pm}_{\mu}(x)(CPT)^{-1}=-W^{\mp}_{\mu}(x)~~~~~~~~~~~~CPTZ^{0}_{\mu}(x)(CPT)^{-1}=-Z^{0}_{\mu}(x)
\end{equation}
\begin{equation}
CPT\bar{\psi}(x)(CPT)^{-1}=\bar{\psi}(x)~~~~~~~~~~~~CPT\psi(x)(CPT)^{-1}=\psi(x)
\end{equation}
So under new $CPT$ transformations, except $W^{\pm}_{\mu}(x)\rightarrow-W^{\mp}_{\mu}(x)$, spinor field and $\gamma_5$ are unchanged while other fields and $\gamma_{\mu}$ change a negative sign. Though the difference between $W^{\pm}_{\mu}(x)$ and $W^{\mp}_{\mu}(x)$, does not affect the calculation results of transition probabilities, it is still unsymmetrical and had better to be improved further to reach a complete symmetry. We can do it by following consideration.
\par
In fact, we can't only regard quantized fields $W^{+}_{\mu}(x)$ and $W^{-}_{\mu}(x)$ as the wave functions of $W^{+}$ and $W^{-}$ particles. In the formulas of weak interaction Hamiltonian, we have
\begin{equation}
W^{+}_{\mu}\bar{\psi}_{\nu}{\gamma}_{\mu}(1+\gamma_5)\psi_l=W^{+}_{\mu}(\bar{\psi}^{(+)}_{\nu}+\bar{\psi}^{(-)}_{\nu})\gamma_{\mu}(1+\gamma_5)(\psi^{(+)}_l+\psi^{(-)}_l)
\end{equation}
\begin{equation}
W^{-}_{\mu}\bar{\psi}_l{\gamma}_{\mu}(1+\gamma_5)\psi_{\nu}=W^{-}_{\mu}(\bar{\psi}^{(+)}_l+\bar{\psi}^{(-)}_l)\gamma_{\mu}(1+\gamma_5)(\psi^{(+)}_{\nu}+\psi^{(-)}_{\nu})
\end{equation}
If $W^{+}_{\mu}(x)$ and $W^{-}_{\mu}(x)$ are only regarded as the wave functions of $W^{+}$ and $W^{-}$ particles, we can not find following four items $W^{+}_{\mu}\bar{\psi}^{(+)}_l\gamma_{\mu}(1+\gamma_5)\psi^{(+)}_{\nu}$, $W^{+}_{\mu}\bar{\psi}^{(-)}_l\gamma_{\mu}(1+\gamma_5)\psi^{(-)}_{\nu}$, $W^{-}_{\mu}\bar{\psi}^{(+)}_{\nu}\gamma_{\mu}(1+\gamma_5)\psi^{(+)}_l$ and $W^{-}_{\mu}\bar{\psi}^{(-)}_{\nu}\gamma_{\mu}(1+\gamma_5)\psi^{(-)}_l$. These four items represent four practical processes, i.e., the creations of $e^{-}$ and $\tilde{\nu}_e$ as well as the annihilations of $e^{+}$ and $\nu_e$ delivered by $W^{+}$ particle, the creations of $e^{+}$ and $\nu_e$ as well as the annihilation of $e^{-}$ and $\tilde{\nu}_e$ delivered by $W^{-}$ particle. All of these four items exist actually. On the other hand, the relation $W^{\pm}_{\mu}(x)=(A^1_{\mu}\pm{i}A^2_{\mu})/\sqrt{2}$ is introduced when we structure wave function $W^{\pm}_{\mu}(x)$ in the electro-weak united theory, meaning that $W^{\pm}_{\mu}(x)$ are the charged complex vector fields. Similar to the charged complex scalar fields, we should have corresponding relations $W^{+}_{\mu}(x)\sim\varphi^{+}(x)$ and $W^{-}_{\mu}(x)\sim\varphi(x)$, so quantized charged vector fields should be written as 
\begin{equation}
W^{+}_{\mu}(x)=W^{+(+)}_{\mu}(x)+W^{+(-)}_{\mu}(x)
\end{equation}
\begin{equation}
W^{-}_{\mu}(x)=W^{-(+)}_{\mu}(x)+W^{-(-)}_{\mu}(x)
\end{equation}
Where $W^{+(+)}_{\mu}(x)$ represents the operator to create a $W^{+}$ particle, $W^{+(-)}_{\mu}(x)$ represents the operator to annihilate a $W^{-}$ particle, $W^{-(+)}_{\mu}(x)$ represents the operator to create a $W^{-}$ particle, $W^{-(-)}_{\mu}(x)$ represents the operator to annihilate a $W^{+}$ particle at space-time point $x$. In this way, similar to complex scalar fields, the time reversal of $W^{\pm}_{\mu}(x)$ fields should also be re-written as
\begin{equation}
TW^{\pm}_{\mu}(x)T^{-1}=-W^{\mp}_{\mu}(\bar{x})
\end{equation}
But the $P,C$ transformations of $W^{\pm}_{\mu}(x)$ are still presented by Eq.(151) and (202). Thus, we have $CPW^{\pm}_{\mu}(x)(CP)^{-1}=W^{\mp}_{\mu}(\bar{x})$ and
\begin{equation}
CPTW^{\pm}_{\mu}(x)(CPT)^{-1}=-W^{\pm}_{\mu}(x)
\end{equation}
\par
Therefore, under new $CPT$ transformations, spinor field and $\gamma_5$ are unchanged, other fields and $\gamma_{\mu}$ change a negative sign. The interaction Hamiltonian constructed by the production of these quantities are either invariable or changing a negative sign. The transition probability is unchanged under new $CPT$ transformations. This result is universal, having nothing to doing with the concrete forms of the Hamiltonians. For the Hamiltonian of present strong, weak and electromagnetic interactions, the new united $CPT$ transformation is at last
\begin{equation}
CPT{\cal{H}}(x)(CPT)^{-1}={\cal{H}}(x)
\end{equation}
It means that the interaction Hamiltonians are completely unchanged. Comparing with the current result
\begin{equation}
CPT{\cal{H}}(x)(CPT)^{-1}={\cal{H}}^{+}(-x)
\end{equation}
it is obvious that Eq.(219) is more symmetric and prefect. Eq.(219) means that the transition probability amplitude is unchanged under $CPT$, while Eq.(220) means that the transition probability density is unchanged under $CPT$, so the $CPT$ invariability in new scheme is a more strict invariability.
\par
Meanwhile, we can obtain $CPTu_s(\vec{p})(CPT)^{-1}=\nu_{-s}(\vec{p})$ and $CPT\nu_s(\vec{p})(CPT)^{-1}=u_{-s}(\vec{p})$ from Eqs.(52), (153), (154), (180) and (181). The results mean that the wave functions of positive and negative particles exchange each other under $CPT$. The result also means that positive particle and negative particle have the same mass when Eq.(219) is considered simultaneously. The results also show that the helicities of spinor particles change a negative sign under $CPT$. For example, the positive particle with left hand helicity is changed into the anti-particle with right hand helicirty. All of these are the same with the current theory, but the descriptions of new scheme are more clear and simple.
\par
On the other hand, by means of Eq.(217), the time reversal of the Hamiltonian of weak interaction between leptons become
\begin{equation}
T{\cal{H}}_w(x){T}^{-1}=-i{g\over{2\sqrt{2}}}[W^{+}_{\mu}(\bar{x})\bar{\psi}_{\nu}(\bar{x})\bar{\gamma}_{\mu}(1+\gamma_5)\psi_l(\bar{x})+W^{-}_{\mu}(\bar{x})\bar{\psi}_l(\bar{x})\bar{\gamma}_{\mu}(1+\gamma_5)\psi_{\nu}(\bar{x})]
\end{equation}
\begin{equation}
T{\cal{H}}_z(x){T}^{-1}=-i{{\sqrt{g^2+g'^2}}\over{4}}{Z}^{0}_{\mu}(\bar{x})[\bar{\psi}_{\nu}(\bar{x})\bar{\gamma}_{\mu}(1+\gamma_5)\psi_{\nu}(\bar{x})+\bar{\psi}(\bar{x})\bar{\gamma}_{\mu}(4{\sin}^2\theta_W-1-\gamma_5)\psi_l(\bar{x})]
\end{equation}
The new $CP$ transformations of the weak interaction Hamiltonian between leptons are
\begin{equation}
CP{\cal{H}}_w(x)(CP)^{-1}=-i{g\over{2\sqrt{2}}}[W^{+}_{\mu}(\bar{x})\bar{\psi}_{\nu}(\bar{x})\bar{\gamma}_{\mu}(1+\gamma_5)\psi_l(\bar{x})+W^{-}_{\mu}(\bar{x})\bar{\psi}_l(\bar{x})\bar{\gamma}_{\mu}(1+\gamma_5)\psi_{\nu}(\bar{x})]
\end{equation}
\begin{equation}
CP{\cal{H}}_z(x)(CP)^{-1}=-i{{\sqrt{g^2+g'^2}}\over{4}}{Z}^{0}_{\mu}(\bar{x})[\bar{\psi}_{\nu}(\bar{x})\bar{\gamma}_{\mu}(1+\gamma_5)\psi_{\nu}(\bar{x})+\bar{\psi}_l(\bar{x})\bar{\gamma}_{\mu}(4\sin^2\theta_W-1-\gamma_5)\psi_l(\bar{x})]
\end{equation}
The $CP$ transformation of quark weak interaction flows is
\begin{equation}
CPJ_{\mu+}(x)(CP)^{-1}=i{g\over{2\sqrt{2}}}\sum^{N}_{j,k=1}{U}_{jk}\bar{d}_k(\bar{x})\bar{\gamma}_{\mu}(1+\gamma_5){u}_j(\bar{x})
\end{equation}
\begin{equation}
CPJ_{\mu-}(x)(CP)^{-1}=i{g\over{2\sqrt{2}}}\sum^{N}_{j,k=1}{U}_{jk}\bar{d}_j(\bar{x})\bar{\gamma}_{\mu}(1+\gamma_5){u}_k(\bar{x})
\end{equation}
Comparing with the current result, the difference is only at $\gamma_{\mu}\rightarrow\bar{\gamma}_{\mu}$. But as mentioned above, this kind of difference does not affect the calculating result of probability densities. Comparing with Eqs.(119)and (120), because $U_{jk}$ is a complex matrix with $U^{\ast}_{jk}\neq{U}_{jk}$ for some matrix elements, we would have $CPJ_{\mu\pm}(x)(CP)^{-1}\neq{J}^{+}_{\mu\pm}(\bar{x})$ in a certain cases, so that $CP{\cal{H}}(x)(CP)^{-1}\neq{\cal{H}}^{+}(\bar{x})$, the symmetry of the $CP$ transformation would be violated. Comparing with Eq.(117) and (118), the situation is also completely the same as $T$ violation, and $T$ violation and $CP$ violation are just complementary according to new scheme. 
\par
Therefore, if regularization and renormalization effects are not considered, new $C,P,T$ scheme is completely the same as the current scheme when we calculate transition probabilities. 
\\
\\
{\bf 6. C,T and CP Violations in High order Renormalization Processes }
\par
As we know that the regularization calculations of some high order processes would cause chirality anomalies in the gauge theory of field. Though it violates gauge invariability, it can be used to solve the forbidden problem of $\pi^{0}\rightarrow{2}\gamma$ decay process caused by the partial conservation of axial vector current. It will be proved according to new scheme that $C,T$ and $CP$ violations would be caused when the regularization and renormalization effects of self-energy and vertex angle in high order perturbation processes are considered, though the united $CPT$ transformation symmetry still holds. The results show that symmetry violations would be common phenomena in the regularization and renormalization processes in the quantum theory of field. 
\par
The high order processes of electromagnetic interaction are taken as examples to show $C,T$ and $CP$ violations below, but the results may be proper for strong and weak interactions. According to renormalization theory, in order to eliminate infinite of electron self-energy, the Hamiltonian of electromagnetic interaction should be revised as
\begin{equation}
{\cal{H}}=-ieN(\bar{\psi}{A}_{\mu}\gamma_{\mu}\psi)-\delta{m}\bar{\psi}\psi
\end{equation}
\\
\\
\\
\\
\\
\\
\\
\\
\\
\par
Fig. 2 High order electron----photon scattering process when mass renormalization is considered
\\
\par
For the high order Compton scattering shown in Fig.2, after mass renormalization is considered, the total transition probability amplitude $S=S_1+S_2+S_3$ can be written as $^{(5)}$
\begin{equation}
S\sim{i}e^2\delta^4(p_1+k_1-p_2-k_2)\bar{u}_s(\vec{p}_2)\epsilon^{\rho}_{\nu}(\vec{k}_2)\gamma_{\nu}{S}^{(2)}_f(p_1-k_1)\gamma_{\mu}\epsilon^{\sigma}_{\mu}(\vec{k}_1){u}_r(\vec{p}_1)
\end{equation}
Here $S^{(2)}_{f}(p)=(m-i\hat{p})/(p^2+m^2)$. Let $p=p_1-k_1$, we have
\begin{equation}
S^{(2)}_{f}(p)=S_f(p)+S_{f}(p)\{\Sigma^{(2)}(p)+i\delta{m}(2\pi)^4\}{S}_{F}(p)
\end{equation}
\begin{equation}
{\Sigma}^{(2)}(p)=e^2(2\pi)^8\int{d}^4{k}\gamma_{\mu}{D}_{f}(k){S}_f(p-k)\gamma_{\mu}
\end{equation}
Before the calculation of regularization, Eq.(228) is unchanged under time reversal with $S^{+}{S}=S^{+}_T{S}_T$. Because $\Sigma^{(2)}(p)$ contains infinite, we have to separate it. By regularization calculation, we get
\begin{equation}
\Sigma^{(2)}(p)=-i\delta{m}(2\pi)^4+BS^{-1}_{f}(p)+S^{-2}_f(p)\Sigma ^{(2)}_f(p)
\end{equation}
Here $B$ is an infinite quantity, but $\Sigma^{(2)}_{F}(p)$ does not contain ultraviolet divergence again with form 
\begin{equation}
\Sigma^{(2)}_{f}(p)={{ie^2}\over{(2\pi)^6}}\int^{1}_{0}{d}x\int^{1}_{0}{d}z{{(1-x)\{(i\hat{p}-m)(1-x)[x-2(1+x)z]+m(1+x)\}}\over{m^2{x}^2+(p^2+m^2)(1-z)xz}}
\end{equation}
The integral can be written simply as
\begin{equation}
\Sigma^{(2)}_{f}(p)=\alpha(p)(p^2+m^2){{i\hat{p}-m}\over{p^2+m^2}}+B(p)=A(p)S_f(p)+B(p)
\end{equation}
Put Eq.(231)into Eq.(229), in light of current method, we obtain
$$\Sigma^{(2)}_{f}(p)=(1+B)S_{F}(p)+\Sigma^{(2)}_f(p)\sim(1+B)\{S_f(p)+\Sigma^{(2)}_{f}(p)\}$$
\begin{equation}
=(1+B)\{[1+A(p)]S_{f}(p)+B(p)\}
\end{equation}
Then we take charge normalization to let $e\rightarrow\sqrt{1+Be}$, Eq.(248) becomes
\begin{equation}
S\sim{i}e^2\bar{u}_s(\vec{p}_2)\epsilon^{\rho}_{\nu}(\vec{k}_2)\gamma_{\nu}\{[1+A(p_1+k_1)]{S}_f(p_1+k_1)+B(p_1+k_1)\}\gamma_{\mu}\epsilon^{\sigma}_{\mu}(\vec{k}_1){u}_r(\vec{p}_1)
\end{equation}
According to the current theory, $S_f(p)$,  $A(p)$ and $B(p)$ are unchanged under time reversal, so the third order renormalization of electron self-energy is invariable under time reversal. It is obvious that the process is also unchanged under $P,C$ transformations.
\par
The regularization and normalization processes of vacuum polarization are discussed below. The total probability amplitude of vacuum polarization containing a second order process and a fourth order process can be written as $S=S_2+S_4$ with $^{(5)}$
\begin{equation}
S=e^2(2\pi)^8\delta^4(p_1-p_3-p_2+p_4)\bar{u}_3(p_3)\gamma_{\mu}{u}_1(p_1){D}^{(2)}_{f,\mu\nu}(p_1-p_3)\bar{u}_4(p_4)\gamma_{\nu}{u}_2(p_2)
\end{equation}
Here $D_f$ is the factor of photon propagation line. Let $k=p_1-p_3$, we have
\begin{equation}
D^{(2)}_{f,\mu\nu}(k)=\delta_{\mu\nu}{D}_f(k)+D_f(k)\Pi^{(2)}_{\mu\nu}(k){D}_f(k)
\end{equation}
\begin{equation}
\Pi^{(2)}_{\mu\nu}(k)=\delta_{\mu\nu}[GD^{-1}_f{k}+\Pi^{(2)}_f(k^2){D}^{-2}_f]
\end{equation}
\begin{equation}
\Pi^{(2)}_{f}(k^2)={{ie^2}\over{3(2\pi)^6}}\int^{1}_{0}{d}y{{y^2(2y-1)(2y-3)}\over{m^2+k^2(1-y)}}
\end{equation}
Here $G$ is an infinite quantity. By charge renormalization to let $e\rightarrow\sqrt{1+Ge}$, Eq.(236) can be written as
\begin{equation}
S\sim{e}^2(2\pi)^8\delta^4(p_1-p_3-p_2+p_4)\bar{u}_3(p_3)\gamma_{\mu}{u}_1(p_1)\delta_{\mu\nu}[1+\Pi^{(2)}_{f}(k_2)]\bar{u}_4(p_4)\gamma_{\nu}{u}_2(p_2)
\end{equation}
Because $\Pi^{(2)}_f(k^2)$ is unchanged under $T$ or $P$  transformations with $\vec{k}\rightarrow-\vec{k}$, so Eq.(240) is invariable. It is obvious that the process is also unchanged under $C$. That is to say that the normalization processes of vacuum polarization are invariable under $C,P,T$ according to the current theory.
\par
At last let us discuss the renormalization of the vertex angle process. The total probability amplitude of the first and third order vertex angle processes can be written as $S=S_1+S_3$ with $^{(5)}$£º
\begin{equation}
S\sim-e\delta^4(p_2-p_1-k_1)\bar{u}_2(p_2)\Gamma^{(2)}_{\mu}(p_1,p_2){u}_1(p_1)a_{\mu}(k_1)
\end{equation}
After regularization calculation is carried out, we have
\begin{equation}
\Gamma^{(2)}_{\mu}(p_1,p_2)=(1+L)\gamma_{\mu}+\Lambda^{(2)}_{f\mu}(p_1,p_2)\sim(1+L)\{\gamma_{\mu}+\Lambda^{(2)}_{f\mu}(p_1,p_2)\}
\end{equation}
In which $L$ is an infinite quantity, but $\Lambda^{(2)}_{f\mu}(p_1,p_2)$ does not contain ultraviolet divergence again.  Then we do charge renormalization to let $e\rightarrow{e}(1+L)$. Thus Eq.(241) becomes
\begin{equation}
S=-e\delta^4(p_2-p_1-k_1)\bar{u}_2(p_2)\{\gamma_{\mu}+\Lambda^{(2)}_{f\mu}(p_1,p_2)\}{u}_1(p_1)a_{\mu}(k_1)
\end{equation}
Here $\Lambda^{(2)}_{f\mu}(p_1,p_2)=G_{\mu}+K_{\mu}$,  $k_1=p_2-p_1$, so we have
\begin{equation}
G_{\mu}={{ie^2}\over{4\pi^4}}\int{d}^4{k}\int^{1}_{0}{d}x\int^{x}_{0}{d}y\{{{R_{\mu}}\over{(k^2+q^2)^3}}-{{\gamma_{\mu}{m}^2(2-2x-x^2/2)}\over{(k^2+m^2{x}^2)^3}}\}
\end{equation}
\begin{equation}
K_{\mu}={{ie^2}\over{4\pi^4}}\int{d}^4{k}\int^{1}_{0}{d}x\int^{x}_{0}{d}y\{{{\gamma_{\mu}}\over{2(k^2+m^2{x}^2)^2}}-{{\gamma_{\mu}}\over{2(k^2+q^2)^2}}\}
\end{equation}
$$R_{\mu}=[\hat{p}_1(1-x)-\hat{k}_1{y}]\gamma_{\mu}(\hat{p}_2-\hat{p}_1{x}-\hat{k}_1{y})-{1\over{2}}{q}^2\gamma_{\mu}-m^2\gamma_{\mu}$$
\begin{equation}
-2im(p_{1\mu}+p_{2\mu}-2p_{1\mu}{x}-2k_{1\mu}{y})
\end{equation}
\begin{equation}
q^2=m^2{x}+p^2_1{x}+k_1(p_1+p_2)y-(k_1{y}+p_1{x})^2
\end{equation}
Eq.(242) can be simplified further. In order to eliminate infrared divergence contained in $\Lambda^{(2)}_{f\mu}(p_1,p_2)$,  we suppose that photon has a small mass $\rho$ before calculation. Then let $\rho\rightarrow{0}$ after calculation. In the low energy problem with $k_1<<m$, the items $k^{n}_1$ with the order $n\geq{3}$  can be omitted. In this way, we can obtain $^{(5)}$
\begin{equation}
\Lambda^{(2)}_{f\mu}(p_1,p_2)=-{{e^2}\over{12\pi^2{m}^2}}[k^2_1\gamma_{\mu}(\ln{m\over{\rho}}-{3\over{8}})+{3\over{4}}m\sigma_{\mu\nu}{k}_{1\nu}]
\end{equation}
Here $\sigma_{\mu\nu}=(\gamma_{\mu}\gamma_{\nu}-\gamma_{\nu}\gamma_{\mu})/2i$. Put it into Eqs.(241) and (242), we get
\begin{equation}
S=-e\delta^4(p_2-p_1-k_1)\bar{u}_2(p_2)\{\gamma_{\mu}-{{e^2}\over{12\pi^2{m}^2}}{k^2_1\gamma_\mu}(\ln{m\over{\rho}}-{3\over{8}})+{3\over{4}}m\sigma_{\mu\nu}{k}_{1\nu}\}u_1(p_1){a}_{\mu}(k_1)
\end{equation}
In the interaction picture of quantum theory of fields, we actually suppose that the electromagnetic field of photon is $A_{\mu}(x)=A_{0\mu}\exp(ik_{1\mu}{x}_{\mu})$, so we have $k_{1\mu}\rightarrow-i\partial_{\mu}{A}_{\mu}$  as well as
\begin{equation}
\sigma_{\mu\nu}{{\partial{A}_{\mu}}\over{\partial{x}_{\nu}}}=-{1\over{2}}\sigma_{\mu\nu}({{\partial{A}_{\mu}}\over{\partial{x}_{\nu}}}-{{\partial{A}_{\nu}}\over{\partial{x}_{\mu}}})=-{1\over{2}}\sigma_{\mu\nu}{F}_{\mu\nu}
\end{equation}
Here $F_{\mu\nu}$ is the tensor of electromagnetic fields. The effective interaction Hamiltonian corresponding to Eq.(249) is 
\begin{equation}
{\cal{H}}(\vec{x},t)=-ie\bar{\psi}(\vec{x},t)\{\hat{A}(\vec{x},t)+{{e^2}\over{12\pi^3{m}^2}}(\ln{m\over{\rho}}-{3\over{8}})\partial^2\hat{A}(\vec{x},t)-{{ie^2}\over{32\pi^2{m}}}\sigma_{\mu\nu}{F}_{\mu\nu}(\vec{x},t)\}\psi(\vec{x},t)
\end{equation}
The third item on the light side of the formula above can be written as $^{(5)}$
\begin{equation}
-{{e^3}\over{16\pi^2{m}}}\bar{\psi}(\vec{x},t)[\vec{\sigma}\cdot\vec{B}(\vec{x},t)-i\vec{\alpha}\cdot\vec{E}(\vec{x},t)]\psi(\vec{x},t)
\end{equation}
Here $\vec{\sigma}$ is Pauli matrices and $\vec{\alpha}$ is the Dirac matrices
\begin{equation}
\sigma_1=\mid^{0~~~1}_{1~~~0}\mid~~~~~~\sigma_2=\mid^{0~~-i}_{i~~~0}\mid~~~~~\sigma_3=\mid^{1~~~0}_{0~~-1}\mid~~~~\vec{\alpha}=\mid^{0~~\vec{\sigma}}_{\vec{\sigma}~~0}\mid
\end{equation}
By using the Lorentz condition $\partial_{\mu}{A}_{\mu}=0$, the electromagnetic potentials can be written as
\begin{equation}
\vec{A}(\vec{x},t)=\vec{A}_0\exp^{(i\vec{k}_1\cdot\vec{x}-\omega{t})}~~~~~~~\varphi(\vec{x},t)={1\over{\omega}}\vec{k}_1\cdot\vec{A}_0\exp^{(i\vec{k}_1\cdot\vec{x}-\omega{t})}={1\over{\omega}}\vec{k}_1\cdot\vec{A}(\vec{x},t)
\end{equation}
By taking the Coulomb gauge to suppose that electromagnetic field is transverse one with $\vec{k}_1\cdot\vec{A}=0$, we have $\varphi=0$ as well as $\vec{B}=i\vec{k}\times\vec{A}$, $E=i\omega\vec{A}$. Put them into Eq.(252), Eq.(251) can be written as 
$${\cal{H}}(\vec{x},t)=-ie\bar{\psi}(\vec{x},t)\{\hat{A}(\vec{x},t)+{{e^2}\over{12\pi^3{m}^2}}(\ln{m\over{\rho}}-{3\over{8}})\partial^2\hat{A}(\vec{x},t)\}\psi(\vec{x},t)$$
\begin{equation}
-{{e^3}\over{16\pi^2{m}}}\bar{\psi}(\vec{x},t)\{i\vec{\sigma}\cdot[\vec{k}\times\vec{A}(\vec{x},t)]+\omega\vec{\alpha}\cdot\vec{A}(\vec{x},t)\}\psi(\vec{x},t)
\end{equation}
According to the current theory, the time reversal of the first and second items in the formula above corresponds to let $t\rightarrow-t$ in $\psi(\vec{x},t)$ and $A_{\mu}(\vec{x},t)$. So they are actually invariable under time reversal when transition probability is calculated. Therefore, we only need to consider the time reversal of the third and fourth items. In light of current theory, let $\sigma_2=i\gamma_1\gamma_3$, we have $\psi(\vec{x},t)\rightarrow\sigma_2\psi(\vec{x},-t)$, $\bar{\psi}(\vec{x},t)\rightarrow\bar{\psi}(\vec{x},-t)\sigma_2$, $\vec{A}(\vec{x},t)\rightarrow-\vec{A}(\vec{x},-t)$ as well as $i\rightarrow-i$, $\vec{k}\rightarrow-\vec{k}$, $\vec{\sigma}\rightarrow\vec{\sigma}^{\ast}$, $\vec{\alpha}\rightarrow\vec{\alpha}^{\ast}$ under time reversal. So the time reversal of the third and fourth items on the light side of Eq.(255) is
\begin{equation}
{{e^3}\over{16\pi^2{m}}}\bar{\psi}(\vec{x},t)\sigma_2\{i\vec{\sigma}^{\ast}\cdot[\vec{k}\times\vec{A}(\vec{x},-t)]+\omega\vec{\alpha}^{\ast}\cdot\vec{A}(\vec{x},-t)\}\sigma_2\psi(\vec{x},-t)
\end{equation}
Because of $\sigma_2\vec{\sigma}^{\ast}\sigma_2=-\vec{\sigma}$,  $\sigma_2\vec{\alpha}^{\ast}\sigma_2=-\vec{\alpha}$, the time reversal of Eq.(275) is
$$T{\cal{H}}(\vec{x},t){T}^{-1}=-ie\bar{\psi}(\vec{x},-t)\{\hat{A}(\vec{x},-t)+{{e^2}\over{12\pi^3{m}^2}}(\ln{m\over{\rho}}-{3\over{8}})\partial^2\hat{A}(\vec{x},-t)\}\psi(\vec{x},-t)$$
\begin{equation}
-{{e^3}\over{16\pi^2{m}}}\bar{\psi}(\vec{x},-t)\{i\vec{\sigma}\cdot[\vec{k}\times\vec{A}(\vec{x},-t)]+\omega\vec{\alpha}\cdot\vec{A}(\vec{x},-t)\}\psi(\vec{x},-t)
\end{equation}
Comparing with Eq.(255), when transition probability is calculated, the result above is invariable under rime reversal.
\par
The $P$ transformation of Eq.(255) is discussed below. According to the current theory, the first and second items are also invariable. When $\vec{k}\rightarrow-\vec{k}$, $\vec{A}(\vec{x},t)\rightarrow-\vec{A}(-\vec{x},t)$ under $P$, the third and fourth items become
\begin{equation}
-{{e^3}\over{16\pi^2{m}}}\bar{\psi}(-\vec{x},t)\gamma_4\{i\vec{\sigma}\cdot[\vec{k}\times\vec{A}(-\vec{x},t)]-\omega\vec{\alpha}\cdot\vec{A}(-\vec{x},t)\}\gamma_4\psi(-\vec{x},t)
\end{equation}
Because of $\gamma_4\vec{\sigma}\gamma_4=\vec{\sigma}$,  $\gamma_4\vec{\alpha}\gamma_4=-\vec{\alpha}$,  the $P$ transformation of Eq.(275) is
$$P{\cal{H}}(\vec{x},t){P}^{-1}=-ie\bar{\psi}(-\vec{x},t)\{\hat{A}(-\vec{x},t)+{{e^2}\over{12\pi^3{m}^2}}(\ln{m\over{\rho}}-{3\over{8}})\partial^2\hat{A}(-\vec{x},t)\}\psi(-\vec{x},t)$$
\begin{equation}
-{{e^3}\over{16\pi^2{m}}}\bar{\psi}(-\vec{x},t)\{i\vec{\sigma}\cdot[\vec{k}\times\vec{A}(-\vec{x},t)]+\omega\vec{\alpha}\cdot\vec{A}(-\vec{x},t)\}\psi(-\vec{x},t)
\end{equation}
Comparing with Eq.(255), when transition probability is calculated, the result is also symmetrical under $P$.
\par
The $C$ transformation is discussed at last. The first and second items are invariable under $C$. According to the current theory, we have $\psi_{\alpha}(x)\rightarrow(\gamma_2)_{\alpha\beta}\psi^{+}_{\beta}(x)$, $\psi^{+}_{\alpha}(x)\rightarrow\psi_{\alpha}(\gamma_2)_{\alpha\beta}$, $\vec{A}(x)\rightarrow-\vec{A}(x)$ under $C$, so the third and fourth items of Eq.(255) become
\begin{equation}
{{e^3}\over{16\pi^2{m}}}\psi_{\alpha}(\vec{x},t)[\gamma_2\gamma_4{i}\vec{\sigma}\cdot(\vec{k}\times\vec{A}(\vec{x},t))\gamma_2+\gamma_2\gamma_4\omega\vec{\alpha}\cdot\vec{A}(\vec{x},t)\gamma_2]_{\alpha\beta}\psi^{+}_{\beta}(\vec{x},t)
\end{equation}
By considering the anti-commutation nature of  fermion's exchange, as well as relations $\gamma_4\vec{\sigma}=\vec{\sigma}\gamma_4$, $\gamma_4\vec{\alpha}=-\vec{\alpha}\gamma_4$,  $\gamma_4\gamma_2=-\gamma_2\gamma_4$, the formula above becomes
\begin{equation}
{{e^3}\over{16\pi^2{m}}}\psi^{+}_{\beta}(\vec{x},t)[\gamma_2{i}\vec{\sigma}\cdot(\vec{k}\times\vec{A}(\vec{x},t))\gamma_2-\gamma_2\omega\vec{\alpha}\cdot\vec{A}(\vec{x},t)\gamma_2]_{\alpha\lambda}(\gamma_4)_{\lambda\beta}\psi_{\alpha}(\vec{x},t)
\end{equation}
Because of $\gamma_2\sigma_1\gamma_2=-\sigma_1$, $\gamma_2\sigma_2\gamma_2=\sigma_2$,  $\gamma_2\sigma_3\gamma_2=-\sigma_3$, $\gamma_2\alpha_1\gamma_2=\alpha_1$, $\gamma_2\alpha_2\gamma_2=-\alpha_2$, $\gamma_2\alpha_3\gamma_2=\alpha_3$, $\sigma^{\tau}_1=\sigma_1$, $\sigma^{\tau}_2=-\sigma_2$, $\sigma^{\tau}_3=\sigma_3$,  $\alpha^{\tau}_1=\alpha_1$, $\alpha^{\tau}_2=-\alpha_2$, $\alpha^{\tau}_3=\alpha_3$, $\gamma^{\tau}_4=\gamma_4$, Eq.(260) becomes
\begin{equation}
-{{e^3}\over{16\pi^2{m}}}\psi^{+}_{\beta}(\vec{x},t)(\gamma_4)_{\beta\lambda}[i\vec{\sigma}\cdot(\vec{k}\times\vec{A}(\vec{x},t))+\omega\vec{\alpha}\cdot\vec{A}(\vec{x},t)]_{\lambda\alpha}\psi_{\alpha}(\vec{x},t)
\end{equation}
So the $C$ transformation of Eq.(255) is
$$C{\cal{H}}(\vec{x},t){C}^{-1}=-ie\bar{\psi}(\vec{x},t)\{\hat{A}(\vec{x},t)+{{e^2}\over{12\pi^3{m}^2}}(\ln{m\over{\rho}}-{3\over{8}})\partial^2\hat{A}(\vec{x},t)\}\psi(\vec{x},t)$$
\begin{equation}
-{{e^3}\over{16\pi^2{m}}}\bar{\psi}(\vec{x},t)\{i\vec{\sigma}\cdot[\vec{k}\times\vec{A}(\vec{x},t)]+\omega\vec{\alpha}\cdot\vec{A}(\vec{x},t)\}\psi(\vec{x},t)
\end{equation}
The interaction Hamiltonian is symmetrical under $C$. Therefore, it is obvious that the renormalization of third order vertex angle process is also symmetrical under the united $CPT$ transformation according to the current theory.
\par
Now let's discuss the symmetry violations in high order processes caused in new scheme when regularization and renormalization are taken into account As shown above, according to new scheme, if the high order processes contain two (or more) Feyman diagrams, among them one contains odd number's fermion inertial lines and another contains even number's fermion inertial lines, time reversal symmetry would be violated by the interference effect. This situation is relative to the mass renormalization of high order Compton scattering. Under new time reversal we have $\bar{\psi}\psi\rightarrow\psi^{\tau}\bar{\psi}^{\tau}=(\bar{\psi}\psi)^{\tau}=\bar{\psi}\psi$, so the additional item of mass renormalization is unchanged. Because the propagation function of spinor field would change a negative sign under time reversal, we have $S_{fT}(p)=-S_f(p)$. So according to new scheme, the time reversal of Eq.(275) is  
\begin{equation}
S_T\sim\bar{u}_s(\vec{p}_1)\epsilon^{\rho}_{\nu}(\vec{k}_1)\gamma_{\nu}\{-[1+A(p_1+k_1)]S_f(p_1+k_1)+B(p_1+k_1)\}\gamma_{\mu}\epsilon^{\sigma}_{\mu}(\vec{k}_2)u_r(\vec{p}_2)
\end{equation}
Comparing with Eq.(235), the interference item would violate time reversal symmetry when transition probability is calculated. 
\par
Similarly, in the high order interaction processes between fermions shown in Fig. 3 (It can be regarded as a part of complex Feyman diagrams.) as well as in more complex diagrams, time reversal symmetries would be violated after mass renormalizations are considered. The fermions can be electron, quark and neutrino and so on. So time reversal symmetry violation may existent commonly in strong, weak and electromagnetic interactions, though concrete calculations should be done for concrete problems. 
\\
\\
\\
\\
\\
\\
\\
\\
\\
\par
Fig. 3 High order fermion scattering process when mass renormalization is considered
\\
\par
On the other hand, under new $C$ transformation, the propagation function of spinor field would also change a negative sign. We have similarly
\begin{equation}
S_C\sim\bar{u}_s(\vec{p}_2)\epsilon^{\rho}_{\nu}(\vec{k}_2)\gamma_{\nu}\{-[1+A(p_1+k_1)]S_f(p_1+k_1)+B(p_1+k_1)\}\gamma_{\mu}\epsilon^{\sigma}_{\mu}(\vec{k}_1)u_r(\vec{p}_1)
\end{equation}
The result also violates the symmetry of $C$ transformation.
\par
Because these symmetry violations appear in the third processes at least, they are very small. No experiments at present can be used to verify them in such high precision. Most of experiments done at present to verify time reversal symmetry are only for low order processes with low precision. It is well known that experiments relative to time reversal are very difficult. However, as shown below, because $T$ violation and $C$ violation are the same according to new scheme when mass renormalization is taken into account, and the experiments for $C$ violation are relatively easy, we can verify $T$ violation indirectly through the experiment of $C$ violation in principle.
\par
Under new $P$ transformation, all $S_f(p)$, $A(p)$ and $B(p)$ are unchanged, so the process is invariable under $P$. In sum, the mass renormalization process of high order Compton scattering violate $C,T$ $CP$ and $PT$ symmetries, but is symmetrical under the united $CT$ and $CPT$ transformations.  It is obvious that under new $C,P,T$ transformations, the renormalization process of vacuum polarization shown in Eq.(240) are unchanged, so it is invariable under united $CPT$ transformation. 
\par
The new transformations of the high order vertex angle renormalization process are discussed below. According to new $T$ transformation, the first and second items in Eq.(255) are also unchanged. We have $\vec{k}\rightarrow-\vec{k}$, $T\vec{A}(x){T}^{-1}=-\vec{A}(\bar{x})$, $T\vec{A}(x){T}^{-1}=-\vec{A}(\bar{x})$, $T\bar{\psi}(x){T}^{-1}=\psi^{\tau}(\bar{x})$, $T\psi(x){T}^{-1}=\bar{\psi}^{\tau}(\bar{x})$, according to new $T$ transformation, so the first and second items in Eq.(255) become
\begin{equation}
-{{e^3}\over{16\pi^2{m}}}\psi^{\tau}_{\alpha}(\bar{x})\{iT\vec{\sigma}T^{-1}\cdot[\vec{k}\times\vec{A}(\bar{x})]-\omega{T}\vec{\alpha}{T}^{-1}\cdot\vec{A}(\bar{x})\}_{\alpha\beta}\bar{\psi}^{\tau}_{\beta}(\bar{x})
\end{equation}
In new scheme we have $T=i\gamma_1\gamma_3\gamma_4$ and get $T\sigma_1{T}^{-1}=\sigma_1$, $T\sigma_2{T}^{-1}=-\sigma_2$, $T\sigma_3{T}^{-1}=\sigma_3$, $T\alpha_1{T}^{-1}=-\alpha_1$, $T\alpha_2{T}^{-1}=\alpha_2$, $T\alpha_3{T}^{-1}=-\alpha_3$, $\sigma^{\tau}_1=\sigma_1$, $\sigma^{\tau}_2=-\sigma_2$, $\sigma^{\tau}_3=\sigma_3$, $\alpha^{\tau}_1=\sigma_1$,  $\alpha^{\tau}_2=-\sigma_2$, $\alpha^{\tau}_3=\sigma_3$. By considering the anti-commutation nature of fermions, the formula above becomes
\begin{equation}
{{e^3}\over{16\pi^2{m}}}\bar{\psi}_{\beta}(\bar{x})\{i\vec{\sigma}\cdot[\vec{k}\times\vec{A}(\bar{x})]+\omega\vec{\alpha}\cdot\vec{A}(\bar{x})\}_{\beta\alpha}\psi_{\alpha}(\bar{x})
\end{equation}
So under new time reversal, Eq.(255) becomes
$$T{\cal{H}}(\vec{x},t){T}^{-1}=-ie\bar{\psi}(\bar{x})[\hat{A}(\bar{x})+{{e^2}\over{12\pi^3{m}^2}}(\ln{m\over{\rho}}-{3\over{8}})\partial^2\hat{A}(\bar{x})]\psi(\bar{x})$$
\begin{equation}
+{{e^3}\over{16\pi^2{m}}}\bar{\psi}(\bar{x})\{i\vec{\sigma}\cdot[\vec{k}\times\vec{A}(\bar{x})]+\omega\vec{\alpha}\cdot\vec{A}(\bar{x})\}\psi(\bar{x})
\end{equation}
The third and fourth items violate time reversal symmetry.
\par
Under new $P$ transformation, the first and second items of Eq.(255) are unchanged. We have $\vec{k}\rightarrow-\vec{k}$, $P\vec{A}(x){P}^{-1}=-\vec{A}(\bar{x})$, $P\bar{\psi}(x)P^{-1}=\bar{\psi}(\bar{x})\gamma_4$, $P\psi(x)P^{-1}=\gamma_4\psi(\bar{x})$ under new scheme, so the third and fourth items become
\begin{equation}
-{{e^3}\over{16\pi^2{m}}}\bar{\psi}(\bar{x})\{i\gamma_4\vec{\sigma}\gamma_4\cdot[\vec{k}\times\vec{A}(\bar{x})]-\omega\gamma_4\vec{\alpha}\gamma_4\cdot\vec{A}(\bar{x})\}\psi(\vec{x})
\end{equation}
Because of $\gamma_4\vec{\sigma}\gamma_4=\vec{\sigma}$, $\gamma_4\vec{\alpha}\gamma_4=-\vec{\alpha}$, under new $P$ transformation, Eq. (255) becomes
$$P{\cal{H}}(\bar{x}){P}^{-1}=-ie\bar{\psi}(\bar{x})[\hat{A}(\bar{x})+{{e^2}\over{12\pi^3{m}^2}}(\ln{m\over{\rho}}-{3\over{8}})\partial^2\hat{A}(\bar{x})]\psi(\bar{x})$$
\begin{equation}
-{{e^3}\over{16\pi^2{m}}}\bar{\psi}(x)\{i\vec{\sigma}\cdot[\vec{k}\times\vec{A}(x)]+\omega\vec{\alpha}\cdot\vec{A}(\bar{x})\}\psi(\bar{x})
\end{equation}
The result is also unchanged under new $P$ transformation.
\par
According to new scheme, the first two items of Eq.(255) are unchanged Under $C$ transformation we have $C\vec{A}(x){C}^{-1}=-\vec{A}(x)$, $C\bar{\psi}(x){C}^{-1}=\psi^{\tau}(x)$, $C\psi(x){C}^{-1}=\bar{\psi}^{\tau}(x)$, so the third and fourth items become
\begin{equation}
{{e^3}\over{16\pi^2{m}}}\psi^{\tau}_{\alpha}(x)[iC\vec{\sigma}{C}^{-1}\cdot(\vec{k}\times\vec{A}(x))+\omega{C}\vec{\alpha}{C}^{-1}\cdot\vec{A}(x)]_{\alpha\beta}\bar{\psi}^{\tau}_{\beta}(x)
\end{equation}
In new scheme $C=i\gamma_2\gamma_4$, we have $C\sigma_1{C}^{-1}=-\sigma_1$, $C\sigma_2{C}^{-1}=\sigma_2$, $C\sigma_3{C}^{-1}=-\sigma_3$, $C\alpha_1{C}^{-1}=-\alpha_1$, $C\alpha_2{C}^{-1}=\alpha_2$, $C\alpha_3{C}^{-1}=-\alpha_3$, $\sigma^{\tau}_1=\sigma_1$, $\sigma^{\tau}_2=-\sigma_2$, $\sigma^{\tau}_3=\sigma_3$, $\alpha^{\tau}_1=\alpha_1$, $\alpha^{\tau}_2=-\alpha_2$, $\alpha^{\tau}_3=\alpha_3$. By considering the anti-commutation nature of fermion's exchange, the formula above becomes
\begin{equation}
{{e^3}\over{16\pi^2{m}}}\bar{\psi}_{\beta}(x)[i\vec{\sigma}\cdot(\vec{k}\times\vec{A}(x))+\omega\vec{\alpha}\cdot\vec{A}(x)]_{\beta\alpha}\psi_{\alpha}(x)
\end{equation}
Therefore, under new $C$ transformation, Eq.(255) becomes
$$C{\cal{H}}(x){C}^{-1}=-ie\bar{\psi}(x)[\hat{A}(x)+{{e^2}\over{12\pi^3{m}^2}}(\ln{m\over{\rho}}-{3\over{8}})\partial^2\hat{A}(x)]\psi(x)$$
\begin{equation}
+{{e^3}\over{16\pi^2{m}}}\{\bar{\psi}(x)\{i\vec{\sigma}\cdot[\vec{k}\times\vec{A}(x)]+\omega\vec{\alpha}\cdot\vec{A}(x)\}\psi(x)\}
\end{equation}
The result is similar to time reversal, the third and fourth items violate $C$ symmetry, so it is also violate $CP$ symmetry. But the united $CT$ and $CPT$ symmetries still hold. 
\par
In sum, according to new scheme, some high order renormalization processes of electromagnetic interaction would violate $C$, $T$, $CP$, $PT$ symmetries, but the united $CT$ and $CPT$ symmetries still hold. This conclusion may be suitable for some high order processes of strong and weak interactions, though concrete calculations are needed. 
\par
By considering Eq.(252) and from Eq.(251), we can write the interaction Hamiltonian of a single as
\begin{equation}
e\varphi-e\vec{\alpha}\cdot\vec{A}+{{e^2}\over{12\pi^2{m}^2}}(\ln{m\over{\rho}}-{3\over{8}})\partial^2\hat{A}-{{e^3}\over{16\pi^2{m}}}\gamma_4(\vec{\sigma\cdot\vec{B}-i\vec{\alpha}\cdot\vec{E}})
\end{equation}
If an electron is in a uniform magnetic field, interaction energy between electron and magnetic field is
\begin{equation}
-e\vec{\alpha}\cdot\vec{A}-{{e^3}\over{16\pi^2{m}}}\gamma_4\vec{\sigma}\cdot\vec{B}=-e\vec{\alpha}\cdot\vec{A}-{{e\alpha}\over{4\pi{m}}}\gamma_4\vec{\sigma}\cdot\vec{B}
\end{equation}
From the formula, we know that an electron's magnetic moment is $\mu=(1+\alpha/2\pi)\mu_0=1.0011614\mu_0$, in which $\alpha\mu_0/2\pi$ is electron's anomalous magnetic moment. So according to new $C$ transformation, the interaction energy between an positive electron and magnetic field, as well as a positive electron's magnetic moment becomes 
\begin{equation}
-e\vec{\alpha}\cdot\vec{A}+{{e\alpha}\over{4\pi{m}}}\gamma_4\vec{\sigma}\cdot\vec{B}~~~~~~~~~\mu=(1-{{\alpha}\over{2\pi}})\mu_0=0.9988386\mu_0
\end{equation}
Here $-\alpha\mu_0/2\pi=-0.0011614\mu_0$ is the anomalous magnetic moment of a positive electron. That is to say, according to new scheme, positive and negative electrons have different magnetic moments. This conclusion can be verified directly by experiments, and the experimental result can be used to prove the correction of new transformation scheme. 
\par
The prediction of $C$ violation shown in Eq.(265) can also be verified directly through experiments, though it needs very high precision for the third order process. After the statistical average of photon's polarization and electron's spin are considered, the differential cross-section of the second order process for the Compton scattering is the so-called Klein-Tamm formula
\begin{equation}
d\Phi={{r^2_0}\over{2}}{{\omega^2_f}\over{\omega^2_i}}({{\omega_f}\over{\omega_i}}+{{\omega_i}\over{\omega_f}}-\cos^2\theta)d\Omega~~~~~~~~~~\omega_f={{\omega_i{m}}\over{m+\omega_i(1-\cos\theta)}}
\end{equation}
\\
\\
\\
\\
\\
\\
\\
\\
\\
\par
 Fig.4 The comparison of theory and experiment for the Compton scattering in the second process
\\
\par
In which $\omega_i$ and $\omega_f$ are the frequencies of initial and final state photons, $\theta$ is the angle between the photon's momentums of initial and final states, $\gamma=k_0/mc^2$, $k_0$ is the momentum of initial photon, $m$ is electron's mass. As shown in Fig.4 $^{(6)}$, the theory coincides with experiment very well. Because the experimental result contains all high order revisions, very fine technique needed to divide the action of mass renormalization of the third process. So the experiment to verify $C$ violation in the high order Compton scattering process would be a great challenge.
\par
The significance of the results above is that it can provide us a method to solve a great problem of so-called reversibility paradox that has puzzled physical circle for along time. According to present understanding, micro-processes are considered reversible under time reversal, but macro-processes controlled by the second law of thermodynamics are always irreversible. We do not know how to solve this contradiction up to now. Though many theories have been advanced, no one is satisfied. As we known that macro-systems are composed of atoms and molecules, atoms and molecules are composed of charged particles. The interactions among charged particles are electromagnetic interaction. According to the discussion above, micro-electromagnetic interaction processes violate time reversal symmetry. So it can be said that the irreversibility of macro-processes originates from the irreversibility of micro-processes actually. 
\par
The result can also be used to explain why our university mainly consists of positive material at present. In the evolution theory of university, this is a fundamental problem. According to observation, the current university mainly consists of positive material, but in the early phase of university, the positive and anti-materials should be the same by the consideration of rationality. In order to evolve from the state that positive and anti-material are symmetric into the state that positive and anti-material are unsymmetrical, big $C$ and $CP$ violations are needed simultaneously. But up to now, only a small $CP$ violation was found in the decay processes of $K^{0}$ and $B^{0}$ mesons. It seams not enough to explain so big asymmetry. According to the paper, $C$ and $CP$ violations also exist commonly in the high order processes of strong and electromagnetic interactions. The result may be useful for us to explain the asymmetry of positive and negative material in university at present.
\\
\\
{\bf References}\\
(1) T.D.Lee, Particle Physics and Introduction to Field Theory, Harwood Academic Publishers, \\
Caption 2.\\
(2) Ying Pengcheng, Essentials of Quantum Theory of Field, Shanghai Science and Technology\\
 Publishing House,125(1986).\\
(3) Luo Changxuan, Introduction to Quantum Theory of Field, Shangxi Normal University \\
Publishing House, 145 (1986).\\
(4) CPLEAR Collaboration(A.Angelopoulos et al.), Phys.Lett.B444, 52(1998). J.Adams et al, \\
Phys.Rev.Lett.80, 4123(1998). P.K.Kabir, Phys.Rev. D2, 540(1970). L.M.Sehgal, \\
M.Wanniger, Phys.Rev. D46, 1035(1992), Phys.Rev.D46, 5209(1992).\\
(5) Zhu Hongyuan, The Quantum Theory of Fields, Science Publishing House, 263£¬272£¬283\\£¨1960£©.\\
(6) Heitler,Walter, The Quantum Theory of Radiation, 220(1954).\\
\end{document}